\documentclass[usenatbib]{mn2e}
\usepackage{cite}
\usepackage{graphicx}
\usepackage{amsmath,amssymb}
\usepackage{verbatim}
\usepackage{subfigure}
\usepackage{pdflscape}
\usepackage{chngcntr}
\usepackage{deluxetable}
\usepackage{color}

\def\lapprox{\lower.4ex\hbox{$\;\buildrel <\over{\scriptstyle\sim}\;$}}
\def\gapprox{\lower.4ex\hbox{$\;\buildrel >\over{\scriptstyle\sim}\;$}}

\title[Non-Detection of Sterile Neutrinos]{Non-Detection of X-Ray Emission From Sterile Neutrinos in Stacked Galaxy Spectra}
\author[Anderson, Churazov, and Bregman]{Michael E. Anderson$^{1}$\thanks{email: michevan@mpa-garching.mpg.de}, Eugene Churazov$^{1,2}$, Joel N. Bregman$^{3}$ \\
$^{1}$Max-Planck Institute for Astrophysics, Garching bei Muenchen, Germany\\
$^{2}$Space Research Institute (IKI), Profsoyuznaya 84/32, Moscow 117997, Russia\\
$^{3}$Department of Astronomy, University of Michigan, Ann Arbor, MI, USA}

\begin{document}

\maketitle

\begin{abstract}
We conduct a comprehensive search for X-ray emission lines from sterile neutrino dark matter, motivated by recent claims of unidentified emission lines in the stacked X-ray spectra of galaxy clusters and the centers of the Milky Way and M31. Since the claimed emission lines lie around 3.5 keV, we focus on galaxies and galaxy groups (masking the central regions), since these objects emit very little radiation above $\sim 2$ keV and offer a clean background against which to detect emission lines. We develop a formalism for maximizing the signal-to-noise of decaying dark matter emission lines by weighing each X-ray event according to the expected dark matter profile. In total, we examine 81 and 89 galaxies with Chandra and XMM-Newton respectively, totaling 15.0 and 14.6 Ms of integration time. We find no significant evidence of any emission lines, placing strong constraints on the mixing angle of sterile neutrinos with masses between 4.8-12.4 keV. In particular, if the 3.57 keV feature from Bulbul et al. (2014) were due to 7.1 keV sterile neutrino emission, we would have detected it at $4.4\sigma$ and $11.8\sigma$ in our two samples. The most conservative estimates of the systematic uncertainties reduce these constraints to $4.4\sigma$ and 7.8$\sigma$, or letting the line energy vary between 3.50 and 3.60 keV reduces these constraints to $2.7\sigma$ and $11.0\sigma$ respectively. Unlike previous constraints, our measurements do not depend on the model of the X-ray background or on the assumed logarithmic slope of the center of the dark matter profile.
\end{abstract}
 
\section{Introduction}

The three known flavors of neutrinos all exhibit left-handed chirality, but the observed oscillations of neutrinos between these flavors have led many theorists to speculate about the possibility of right-handed neutrinos as well (e.g. \citealt{Majorana1937}, \citealt{Minkowski1977}, \citealt{Mohapatra1980}). These right-handed neutrinos would lack the weak coupling of their left-handed counterparts and have therefore been termed ''sterile neutrinos''. 

Sterile neutrinos offer natural solutions to several astrophysical problems. In addition to explaining neutrino mixing, their existence would also explain the matter-antimatter asymmetry in the Universe (\citealt{Akhmedov1998}, \citealt{Shaposhnikov2008}). Moreover, they are non-baryonic and lack electromagnetic coupling, and therefore represent a possible dark matter candidate (\citealt{Dodelson1994}, \citealt{Shi1999}, \citealt{Abazajian2001}, \citealt{Dolgov2002}). In the "neutrino minimal standard model" ($\nu$MSM, \citealt{Asaka2005}), there are three flavors of sterile neutrinos, of which the least massive has a mass $\gapprox 400$ eV, which could plausibly account for the bulk of the dark matter in the Universe. 

While sterile neutrinos lack permitted electromagnetic interactions, they can decay spontaneously. The channel which is considered most likely to be detectable is spontaneous decay into a left-handed neutrino and a photon. The photon has an energy equal to half the mass of the sterile neutrino. For sterile neutrinos with masses in the keV range, this corresponds to an X-ray photon. The decay rate $\Gamma$ is determined by the mixing angle $\theta_1$ of the sterile neutrino according to the following relation (\citealt{Pal1982}; \citealt{Barger1995}; \citealt{Boyarsky2009}):

\begin{equation} \Gamma = 1.38\times10^{-22} \text{ sin}^2(2\theta) \left(\frac{m_S}{\text{1 keV}}\right)^5 \text{ s}^{-1} \end{equation}

\noindent where $m_S$ is the mass of the sterile neutrino. The mixing angle is traditionally denoted as sin$^2 (2\theta)$ instead of $\theta_1$ (in any plausible model $\theta_1 \ll 1$), and we will use that notation in the rest of this work as well. \citet{Boyarsky2009} have compiled a number of constraints on $m_S$ and sin$^2(2\theta)$ from the physics and astrophysics literature.

Within the last year, two teams have provided potential observational evidence for the existence of such a sterile neutrino. The first team (\citealt{Bulbul2014}, hereafter Bu14) examined the X-ray spectra of the intracluster media in 73 galaxy clusters with redshifts between 0.01 and 0.35 using the XMM-Newton telescope. After shifting the spectra to the rest-frame, they stacked them together, and also examined several sub-samples of these clusters. The effective exposure times of these stacks are $~2$ megaseconds on the PN instrument and $~6$ megaseconds on the MOS instruments, which are roughly equivalent to continuous integrations of 1 and 2.5 months duration, respectively.

After carefully modeling the continuum and any known atomic transitions in the spectra, Bu14 find a weak emission line in these spectra which corresponds to no known atomic transition. The exact energy of this line and its flux vary slightly depending on the detector and the subsample of galaxy clusters under consideration. The XMM-MOS spectrum of the full stack of 73 clusters shows the line at $3.57\pm0.02$ keV  with a flux of $4.0\pm0.8 \times 10^{-6}$ photons cm$^{-2}$ s$^{-1}$ ($1\sigma$ confidence intervals). Including the line in their model improves the $\chi^2$ of the fit by 22.8 at a cost of 2 degrees of freedom, which corresponds to a $4.4\sigma$ detection. The look-elsewhere effect\footnote{The energy of the line is unknown {\it a priori} so the significance must be degraded based on the number of independent energies at which they searched for the line.} reduces the significance somewhat, but Bu14 claim their detection still exceeds $3\sigma$ significance.

To show that this result is not instrument-dependent, Bu14 also examine the spectra of two galaxy clusters observed with the Chandra observatory. They find the same unidentified line in observations of Perseus with the ACIS-S3 chip at an energy of $3.56\pm0.02$ keV, although the significance is lower ($\Delta \chi^2 = 11.8$ for 2 d.o.f.). They also find a similar result for Chandra ACIS-I observations of Perseus, albeit at even lower significance. In the Virgo cluster, this line is not detected with Chandra, but the upper limit is roughly consistent with the expected flux from the line based on their detections in other systems.

The second study (\citealt{Boyarsky2014a} hereafter Bo14) also examines XMM-Newton observations of the Perseus cluster. Their model for the intracluster plasma emission is somewhat simpler than the model adopted by Bu14, but the results are similar: they find an unidentified line at $3.50\pm0.04$ keV in the stacked MOS spectra and at $3.46\pm0.04$ in the stacked PN spectra. The significance of these detections is a bit lower than Bu14, with $\Delta \chi^2 = 9.1$ and $8.0$ respectively (for 2 d.o.f.). However, they are also able to show that the strength of the line decreases with projected radius from the center of Perseus, and that the profile appears slightly ($\sim 1.5 \sigma$) more consistent with the expected dark matter profile than with the intracluster gas profile. 

Bo14 perform a similar analysis for XMM-Newton observations of M31. They find a line at $3.53\pm0.03$ keV with $\Delta \chi^2 = 13.0$ for 2 d.o.f., and show that this line also disappears at larger projected radii. A combined analysis of M31 and Perseus shows the line at $3.52\pm0.02$ keV, with $\Delta \chi^2 = 25.9$ for 3 d.o.f., a $4.4\sigma$ detection before accounting for the look-elsewhere effect. 

The inclusion of M31 is an important contribution, since it is the only object in these works which is not a galaxy cluster. Galaxies and galaxy clusters are both dark-matter-dominated systems, a galaxy cluster is suffused with a hot ($kT > 2$ keV) intracluster medium (ICM) which contains most of the baryonic mass associated with the cluster. This plasma produces a bright X-ray continuum (primarily, but not exclusively, thermal bremsstrahlung radiation) which can span the entire observable bands of Chandra and XMM-Newton even for intermediate temperature clusters like Perseus. The ICM is also substantially metal-enriched, so careful modeling of all known atomic transitions is necessary in order to distinguish potential emission lines from non-atomic sources like sterile neutrinos. Even worse, while single-temperature collisional ionization equilibrium models generally give adequate fits to observed ICM spectra, in detail the ICM is multiphase (e.g. Peterson et al. 2003) and contains shocks (e.g. Fabian et al. 2006), so non-equilibrium effects must also be included in a model of X-ray emission from the ICM.

The X-ray spectra of galaxies have none of these complications. Massive ellipticals (\citealt{Forman1985}, \citealt{Fabbiano1989}) and at least some massive spirals (\citealt{Anderson2011}, \citealt{Dai2012}, \citealt{Bogdan2013}) are surrounded by hot gaseous halos, but the temperatures of these halos are  $< 1$ keV, so their emission above $\sim 2$ keV is negligible. There are a few other sources of harder X-rays in galaxies, but the majority of this emission can be localized to individual point sources which can be masked. Recent papers by \citet{RiemerSorensen2014}, \citet{Jeltema2014}, and \citet{Boyarsky2014b} make use of this point to study sterile neutrino emission in galaxies, but these studies all examine galactic nuclei. These regions are often filled with hard X-ray emission and therefore lack some of the advantages offered by the outskirts of galaxies. These issues are discussed further in section 5.

In this paper, we focus on galaxies as potential sources of X-ray line emission from sterile neutrinos. We present a stacking methodology which is optimized for the detection of decaying dark matter emission from the outskirts of galaxies (section 2). We apply this methodology to a sample of 81 galaxies observed with Chandra and a sample of 89 galaxies observed with XMM-Newton (section 3), with total integration times of 15.0 and 14.6 megaseconds respectively. Stacking the spectra of these galaxies, we perform deep searches for unidentified X-ray lines in the range of 2.4 - 6.2 keV (section 4). We do not detect any such lines. We present our upper limits and discuss the implications of these results in section 5. In this paper we generally refer to ``sterile neutrino'' lines, but note that our methodology and results are applicable for any decaying dark matter - type emission.

\section{Methods}

\subsection{Generating Spectra}

\subsubsection{Chandra data reduction}

We use the procedure outlined in \citet{Vikhlinin2005} in order to process the level 1 event files, generate exposure-corrected spectra over the full field of view of each observation, and combine these spectra into a single stacked spectrum for each galaxy. 

We begin with the level 1 event files, and perform automated processing to remove flares and periods of especially high background. We the combine the level 2 event files to produce a broad-band (0.5-7.0 keV) exposure map and vignetting-corrected image. We use these merged images in order to identify point sources to exclude from the spectral extraction regions. In order to identify point sources, we first run the \verb"wavdetect" procedure from the Chandra Interactive Analysis of Observations (CIAO) version 4.4.1, supplying the merged broad-band image and exposure map, and using wavelet scales of powers of two ranging from $2^0$ to $2^4$ pixels. The sensitivity threshold is $10^{-6}$, which can be expected to flag a number of false positives in the large images we consider here. As a second step, we perform aperture photometry on each potential point source, using the individual images, exposure maps, and psf maps from every observation which includes that point source, and taking local backgrounds from locations near the point source in the observation. Any point source which is significant at $\ge 5\sigma$ is masked, out to a radius of 1.5 times the 90\% encircled counts radius of the psf at 3.75 keV. We also visually examined the broad-band images to ensure this technique does an adequate job of identifying and masking the bright point sources, and mask additional sources by hand when necessary.

\subsubsection{XMM data reduction}

We use a modified version of the same procedure to process the XMM PPS data products. We calculate good time intervals by scaling the 9.0-12.0 keV lightcurve, and then generate a merged, vignetting-corrected broad-band image from the de-flared event files. We run  \verb"wavdetect" on these images, using default psf size of 8'', in order to detect point sources. We then manually adjust individual source ellipses as necessary, primarily to increase the size of the mask around the center of extremely X-ray bright sources such as NGC 2992.

\subsubsection{Generating Dark-Matter-weighted spectra}

When we extract the X-ray spectrum from the un-masked regions covered by the event files, we weight each event (as a function of its energy and position on the detector) in order to correct for vignetting, using a power-law spectrum ($\Gamma = 2$) as a reference model. This produces an image with spatially uniform effective area equal to an on-axis ACIS-I observation (Chandra) or MOS observation (XMM-Newton). This weighting scheme yields non-integer photon counts, so we must use Gaussian statistics instead of Poisson statistics, but with $\sim15$ MS of integration time in our stacked spectra, we have enough counts for Gaussian statistics to be an appropriate approximation. 

In order to optimize the search for emission from sterile neutrinos, we also weight each event by the expected dark matter column at that location. To do this, we need an estimate of the virial mass and virial radius of each galaxy. Our estimates are fairly crude, but probably correct to within a factor of two for most individual galaxies, and unlikely to be biased significantly in either direction. We compute the absolute K-band magnitude of the galaxy, using the 2MASS observed total K-band magnitude and the average redshift-independent distance to the galaxy (both taken from the NASA Extragalactic Database). We then assume a K-band M/L ratio of 0.5 to infer the stellar mass, noting that in the K-band this ratio is not very dependent on galaxy mass or morphology \citep{Bell2001}. We then infer the halo mass by interpolating the stellar-to-halo-mass relation of \citet{Moster2010} (in Appendix A we examine the effect of using the \citealt{Behroozi2010} relation instead). Finally, we take the virial radius to be

\begin{equation} R_{\text{vir}} = \left( \frac{M_{\text{halo}}}{200 \times \frac{4}{3} \pi \rho_c} \right)^{\frac{1}{3}} \end{equation} 

\noindent where $\rho_c$ is the critical density of the Universe, $9.1\times10^{-30}$ g cm$^{-3}$, and we neglect the redshift dependence of this parameter since all the galaxies we examine are at very low redshift $(z < 0.03)$. Inferred halo masses and virial radii for each of our galaxies are listed in Table 1. 

Generally, this method gives reasonable results, but in a handful of cases either the average redshift-independent distance for a galaxy or the stellar mass inferred form the total 2MASS $K_S$-band flux disagree significantly with accepted measurements. In five cases, this leads to implausibly large halo masses, so for these five cases we lowered the distance and/or halo masses in order to be conservative. These five cases are NGC 1316, NGC 1961, NGC 6482, NGC 6753, and NGC 6876.

\subsection{Constraining the Mixing Angle}

Here we present our formalism for weighting each photon by the expected dark matter density. We follow Bu14 and Bo14 in defining the parameter of interest to be the mixing angle sin$^2(2\theta)$ of the sterile neutrino dark matter. Since sterile neutrino decay is a spontaneous process, its emissivity will depend on the column density $\Sigma_{\text{DM}}$ of dark matter within our beam (of angular size $\Omega$). For this study, the dark matter in the beam is associated with a galaxy at distance $D$ from Earth. We start from equation (1) and multiply both sides by the number of sterile neutrinos in the beam, assuming all the matter is in dark matter\footnote{This neglects the contribution of baryons to the total galaxy mass, but the Cosmological baryon fraction is only 0.17 \citep{Planck2013}, and most galaxies seem to have baryon fractions significantly lower than this \citep{Anderson2010}, suggesting these systems are truly dominated by dark matter.}  

\begin{equation}
\begin{aligned} L_{s} &= 1.38\times10^{-29} \text{ photon s}^{-1}\\
 &\times \left(\frac{\text{sin}^2 \text{ } 2\theta}{10^{-7}}\right) \left(\frac{m_S}{1 \text{ keV}}\right)^4 \left(\frac{M_{\text{DM}}}{1 \text{ keV}}\right) \end{aligned}
 \end{equation}

\noindent where $L_s$ is the number of photons generated by the decay of sterile neutrinos within the field of view of the beam. Each photon has an energy of $m_S / 2$ keV. Converting to specific intensity,

\begin{equation}
\begin{aligned} I_s &= 1.38\times10^{-29} \text{ photon s}^{-1} \text{ cm}^{-2} \text{ sterad}^{-1}\\
&\times \left(\frac{\text{sin}^2 \text{ } 2\theta}{10^{-7}}\right) \left(\frac{m_S}{1 \text{ keV}}\right)^4 \left(\frac{M_{\text{DM}}}{1 \text{ keV}}\right) \frac{1}{4 \pi D^2 \Omega } \end{aligned}
\end{equation}

\noindent Noting that $M_{\text{DM}}  = \Sigma_{DM} \Omega D^2$, this becomes

\begin{equation} \begin{aligned}I_S &= 1.45\times10^{-11} \text{ photon s}^{-1} \text{ cm}^{-2} \text{ arcsec}^{-2} \\
&\times\left(\frac{\text{sin}^2 \text{ } 2\theta}{10^{-7}}\right) \left(\frac{m_S}{1 \text{ keV}}\right)^4 \frac{\Sigma_{\text{DM}}}{\text{ g cm}^{-2}} \end{aligned}\end{equation}

If we have a good ${\it a}$ ${\it priori}$ expectation for the distribution of dark matter within the virial radius, we can optimize the S/N of this estimate by dividing the image into bins (pixels). We use square bins with sides of length 1'' for the Chandra observations and 2''  for the XMM-Newton observations. We define an estimator $\alpha$ as 

\begin{equation}\alpha \equiv 10^{7} \text{ sin}^2 (2\theta) \left(\frac{m_S}{1 \text{ keV}}\right)^4 \end{equation}

\noindent Then, in each pixel $i$,

\begin{equation}
\alpha_i = \frac{I_i}{M_i} 
\end{equation}

\noindent Where we have used $M = 1.45\times10^{-11} \times  \Sigma_{\text{DM}} \text{ photon s}^{-1} \text{ cm}^{-2} \text{ arcsec}^{-2}$ to denote the model prediction. If we neglect uncertainties in the model, then the uncertainty $\sigma_{\alpha_i}$ in the estimate of $\alpha$ in pixel $i$ is given by

\begin{equation}
\sigma_{\alpha_i} = \frac{\sigma_{I_i}}{M_i} 
\end{equation}

\noindent Where $\sigma_{I_i}$ is the measurement uncertainty for $I$ within that pixel. The parameter we seek to compute is the uncertainty-weighted mean of $\alpha$, which is defined as

\begin{equation}
\left< \alpha \right> \equiv \frac{\Sigma_i{{\alpha_i} / \sigma_{\alpha_i}^2}}{\Sigma_i{1 / \sigma_{\alpha_i}^2}}
\end{equation}

\noindent This gives us the optimal estimate for $\alpha$ which can be derived from the data.

We assume the dark matter follows an NFW profile \citep{Navarro1997}. We note that NFW profiles predict a cusp in the center of the halo, so weighting by $\Sigma_{DM}$ would cause the central pixels to dominate the total expected signal. Moreover, while the NFW profile is generally found to be acceptable (\citealt{Gavazzi2007}, \citealt{Newman2013}), the behavior at the center of the halo (particularly the central logarithmic slope) is controversial (e.g. \citealt{deblok2010} and references therein). Finally, the centers of galaxies are also often hosts of active nuclei which could contaminate a potential signal. For all these reasons, we exclude the pixels within the central $0.01R_{\text{vir}}$ of the galaxy from our analysis (for M31 we include pixels down to radii of $0.005 R_{\text{vir}}$, or 0.94 kpc, which is roughly the size of the bulge). 

An NFW profile has two shape parameters -- $R_{\text{vir}}$ and the concentration $c$ -- as well as a total mass $M_{\text{vir}}$ which is tied to $R_{\text{vir}}$. We have already defined $R_{\text{vir}}$ and $M_{\text{vir}}$ (section 2.1.3). For $c$, we use the $M$-$c$ relation of \citet{Prada2012}, which gives $c$ as a function of $M_{\text{vir}}$ and $z$ (we take $z=0$). We have explored the $M$-$c$ relation of \citet{Zhao2009} as well, and find that the exact form of the relation is not very significant (Appendix A). In Appendix A we also show that errors in our DM model do not have a very significant effect on the results. The projected DM density at impact parameter $b$ in an NFW profile is given by (see also \citealt{Bartelmann1996} for an alternate expression)

\begin{equation} \begin{aligned}
\Sigma_{DM}(b) &= \frac{M_{\text{vir}}}{2 \pi R_{\text{vir}}^2} \frac{c^2}{\text{log}(1+c)-c/(1+c)} \times f(x)
\end{aligned}
\end{equation}

\noindent where $x \equiv cb/R_{\text{vir}}$, 

\begin{equation}
f(x)=  \left\{
     \begin{array}{rr}
       1/3 & , x=1\\
	\left(1 - \frac{g(x)}{\sqrt{\left|x^2 -1\right|}} \right) \frac{1}{x^2 -1} 
 & , x \neq1\\
     \end{array}
   \right.
\end{equation}

\noindent and 
\begin{equation}
g(x) =  \left\{
     \begin{array}{rr}
     \text{arccos}(x^{-1}) & , x > 1\\
 \text{arccosh}(x^{-1}) & , x < 1
 \end{array}
 \right.
\end{equation}

\section{Sample Selection}

We searched the HyperLEDA catalog for all galaxies whose virial radii (estimated by scaling the optical photometry) subtend more than an arcminute in the sky. We then cross-matched this list of galaxies with the Chandra and XMM archives, selecting any of these galaxies which have publicly available observations totalling 50 ks or longer. We discarded any galaxies which are members of galaxy clusters or groups with $kT \gapprox 1$ keV, in order to minimize the X-ray background. We also discarded a few observations (primarily Chandra observations taken during periods early in the lifetime of the telescope), for which the calibrations are less certain. 

Several obsids contained multiple galaxies which fit our criteria. In these cases, we selected the most luminous galaxy in the field as our target, and discarded the others. This means that, for these systems, we underestimate the total projected column density of dark matter (since we neglect the other galaxies in the field). This is the most conservative way to model these fields, and leads to the most robust constraint on the emission from sterile neutrinos around the primary galaxy. We also note that the virial radii of NGC6861 and NGC6868 overlap, but the X-ray observations of these galaxies do not overlap. Each obsid we consider is matched to one and only one galaxy. 

The final list of targets is presented below in Table C1. The columns $t_{\text{CXO}}$ and $t_{\text{XMM}}$ are the total integration times with these telescopes for each galaxy; the flaring-corrected times are generally lower, but we apply the flaring correction to each chip (Chandra) or MOS detector (XMM) individually, so it is not straightforward to define a single flaring-corrected exposure time for each galaxy. The total integration times for our samples are 15.0 Ms for Chandra and 14.6 Ms for XMM-Newton.

\section{Results}

In Figure 1 we present representative Chandra and XMM-Newton spectra. The spectra are displayed in flux units of the estimator $\alpha$, which is related to the mixing angle sin$^2(2\theta)$ (see section 2.2). 

Each spectrum has been shifted to the rest-frame using its measured recessional velocity (in practice, since our sample consists of galaxies within 100 Mpc, this shift is insignificant). No background subtraction has been applied, and all of these spectra are background-dominated above $\sim 2$ keV. The visible emission lines are instrumental features and are identified in the plot. 

\begin{figure}
\begin{center}
\includegraphics[width=9cm]{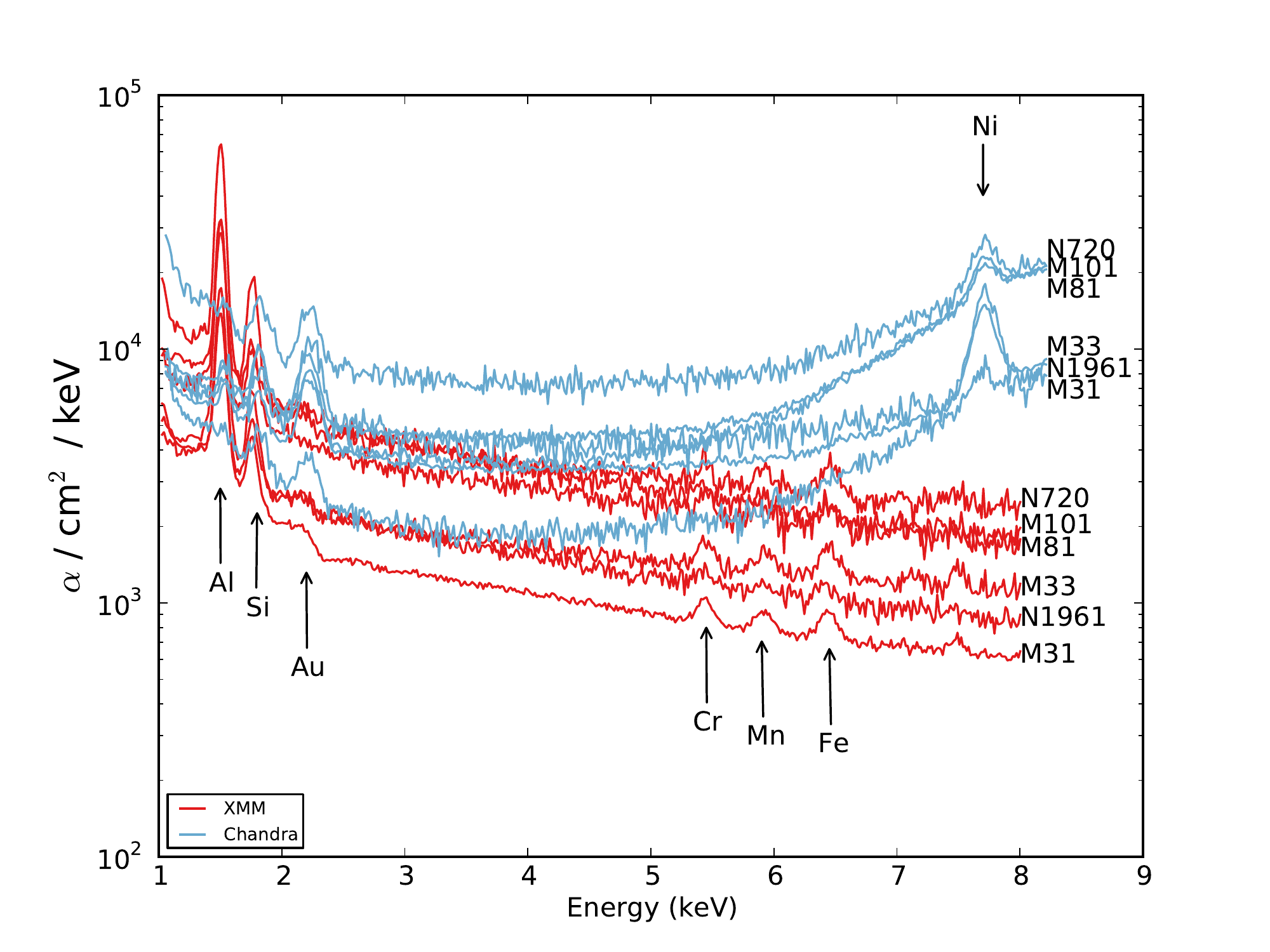}
\end{center}
\vspace{-0.5cm}
\caption{Representative spectra of five galaxies, weighted by projected dark matter column density, displayed in units of our estimator $\alpha$ which is defined in the text. Above $\sim 2$ keV, these spectra are dominated by the instrumental background. Prominent instrumental features are labeled, and for Chandra spectra the strength of the instrumental uptick at high energies depends on the ratio of FI to BI chips in each individual observation, as well as the activity of the Sun during the observations. }
\end{figure}

We then compute the weighted average of each of these galaxies, using the formalism described in section 2.2. The weight depends on the measurement uncertainty in each spectrum and the total estimated exposure-weighted dark matter column within each observation. We use a canned rmf file tied to the ACIS-I FI chips for Chandra and to the MOS1 detectors for XMM-Newton. We stack the weighted, de-redshifted arf files in the same way as we stack the spectra.

In Figure 2, we show the distribution of expected statistical strengths of the sterile neutrino emission from each galaxy. This is proportional to the expected S/N value for the emission from each galaxy, which it can be shown is proportional to the weighted sum over each pixel of the quantity $\sqrt{\Sigma_{\text{DM}} \Omega T/\alpha}$, where $T$ is the exposure time per pixel and $\alpha$ is our estimator defined in equation 6. We plot this ratio for each galaxy in Figure 2, using the flux measured over the 3-4 keV band (note that $\alpha$ is energy-dependent). 

The distribution spans roughly an order of magnitude, though the five most promising individual targets are all XMM-Newton sources. These five targets all have a combination of long exposure times, low distances (so their dark matter halos fill the entire XMM field of view) and low flux in the 3-4 keV band. Their contribution to the total stacked signal should be proportional to their statistical strength; these five galaxies collectively comprise 16\% of the expected signal (and 6\% of the sample), so they have an important effect on the result but do not dominate it. 

\begin{figure}
\begin{center}
\includegraphics[width=9cm]{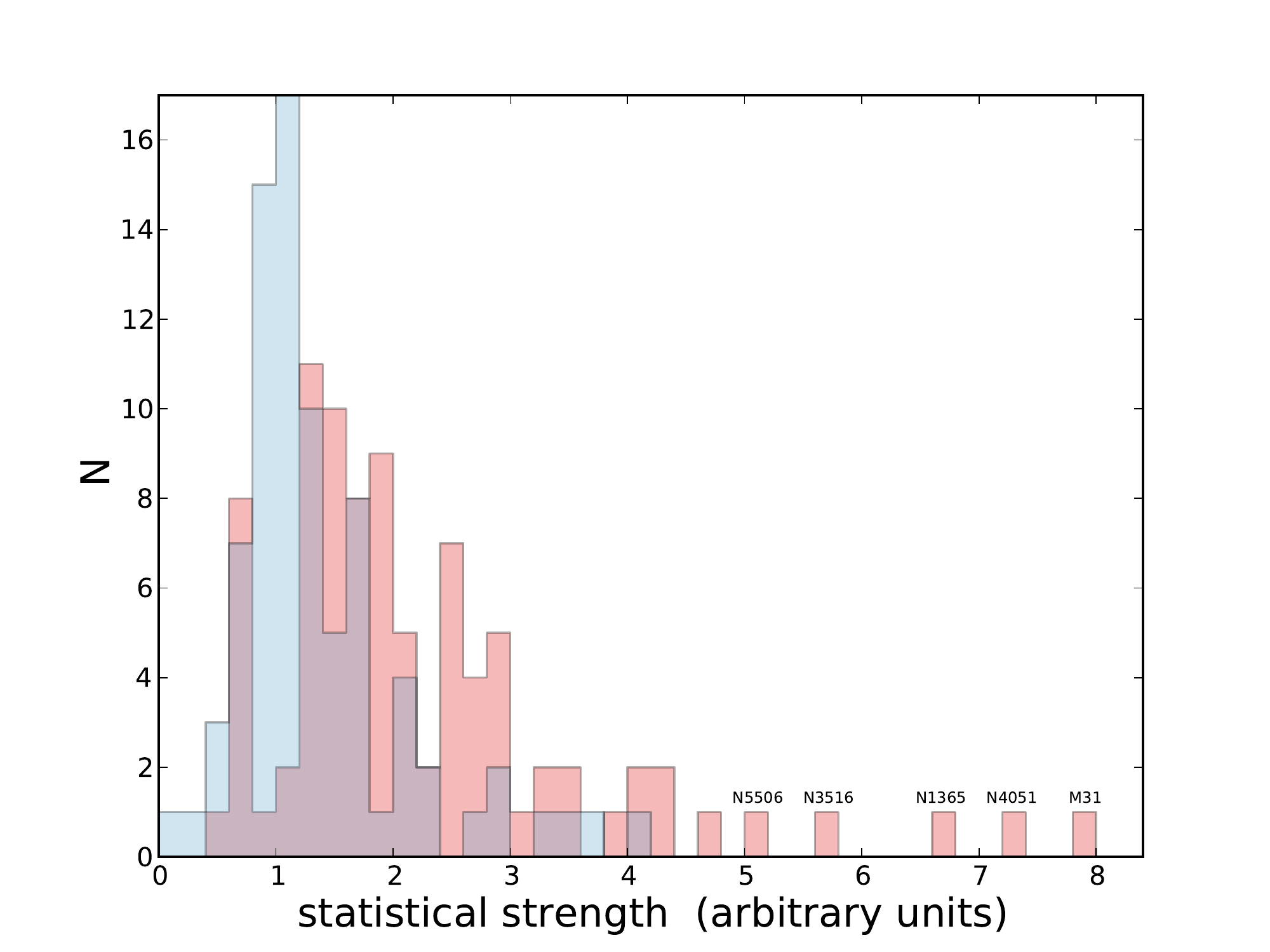}
\end{center}
\vspace{-0.5cm}
\caption{ Distribution for our samples of 81 Chandra (blue) and 89 XMM-Newton (red) galaxies of the expected statistical strength of the sterile neutrino emission. This quantity is proportional to the expected signal-to-noise of an emission line (the constant of proportionality depends on the mixing angle and the exact energy of the line). Our galaxies span about a range of about an order of magnitude in expected signal-to-noise, and the XMM-Newton sample extends to stronger values than the Chandra sample. The five galaxies with the strongest expected signals are indicated; they collectively comprise 16\% of the expected XMM-Newton signal.  }
\end{figure}

\subsection{Modeling the X-ray Spectrum}

A major advantage of our approach is that galaxies have very little X-ray emission above $\sim 2$ keV, especially after masking bright point sources. The dominant contribution to these spectra is the instrumental background. A common approach is to subtract the instrumental background, scaling it from canned observations while the telescope is stowed. However, the stowed background files are comprised of only $\lapprox 1$ Ms of integration time, which is much lower than the $\sim 15$ Ms of integration time of our stacked spectra. The uncertainties in the stowed backgrounds are therefore much larger than the uncertainties in our stacked spectra, so if we subtract scaled stowed backgrounds we increase our total uncertainty budget significantly. An alternative would be to construct our own model of the spectral and spatial shape of the instrumental background, although this procedure would introduce additional assumptions and does not offer significantly tighter constraints than we are already able to derive.

Instead of modeling or subtracting the instrumental background, we fit it with a smoothing spline. We analyze the spectra using a combination of XSPEC v. 12.8.1 \citep{Arnaud1996}, PyXspec v.1.0.3, and the \verb"UnivariateSpline" class included in SciPy v. 0.13.3. This implementation uses error-weighted B-splines; we increase the degree of the polynomial (up to fourth degree) until we get an acceptable fit to the spectrum.  We do not allow the spline to generate any new knots, so the two knots are fixed to the edges of the spectral extraction region. The spline allows us to fit the large-scale variations in the instrumental background visible in Fig. 3 (particularly the turn-up at large energies), while remaining sufficiently inflexible that it allows for lines to be distinguished from the continuum and their flux recovered correctly (Appendix B). 

We select regions of the spectrum in between prominent instrumental lines, in order to get the cleanest measurement of the continuum (in particular, we avoid Au at the low end, Cr at the high end for XMM-Newton MOS, and Ni at the high end for Chandra ACIS). We fit to the 2.6-5.2 keV band for XMM-Newton and to the 2.4-6.2 band for Chandra. The spline fits to each of our stacked spectra are shown in Table 1. 

We also fit the same spectra with a model containing both a smoothing spline and a \verb"Gaussian" component at $3.57$ keV with zero width. In the "free line" case, we let the normalization of the line float (across positive and negative numbers), and in the "fixed line" case we freeze the normalization at a value corresponding to a mixing angle of $7\times10^{-11}$ (the best-fit value from Bu14). In both cases, the spline is fit to the difference between the data and the line profile (which is folded through the instrumental response). The smoothing is error-weighted, using the sum of errors on the data and noise in the line, added in quadrature.

\begin{table*}
\begin{minipage}{146mm}
\begin{tabular}{lccccccc}
\hline
name & C0 & C1 & C2 & C3 & C4 & $\chi^2$ & d.o.f.\\
\hline
XMM-Newton no line & $1.57\times10^5$ & $1.35\times10^5$ & $1.24\times10^5$ & $1.13\times10^5$ & $1.03\times10^5$ & 137.5 & 167 \\
XMM-Newton fixed line & $1.57\times10^5$ & $1.34\times10^5$ & $1.23\times 10^5$ & $1.14\times10^5$ & $1.02\times10^5$ & 305.8 & 167\\
XMM-Newton free line & $1.57\times10^5$ & $1.36\times10^5$ & $1.25\times10^5$ & $1.13\times10^5$ & $1.03\times10^5$ & 131.6 & 166\\
Chandra no line & $3.10\times10^5$ & $2.29\times10^5$ & $2.94\times10^5$ & $2.30\times10^5$ & $3.70\times10^5$ & $231.1$ & 254\\
Chandra fixed line & $3.10\times10^5$ & $2.31\times10^5$ & $2.94\times10^5$ & $2.29\times10^5$ & $3.70\times10^5$ & 250.8 & 254\\
Chandra free line & $3.16\times10^5$ & $2.29\times10^5$ & $2.94\times10^5$ & $2.29\times10^5$ & $3.67\times10^5$ & $231.1$ & 253\\
\end{tabular}
\caption{Spectral fits to the stacked spectra shown in Figure 3. Each fit uses a B-Spline which is described by two knots at the edges of the spectral band under consideration and five parameters $C0$-$C4$ corresponding to the coefficients of a fourth-degree polynomial. The ''with line'' fit also includes a Gaussian component in the model (folded through the instrumental response) with parameters corresponding to the best-fit value from Bu14 (see text), and the "free line" fit allows this Gaussian component to have any arbitrary normalization (positive or negative). }
\end{minipage}
\end{table*}

In both cases, the "no line" model is much more strongly favored than the "free line" model. Using an F-test, we find that the "free line" model always has a best-fit mixing angle which is consistent with zero within $3\sigma$. On the other hand, comparing the "fixed line" to the "free line" model, the former is ruled out at $4.4\sigma$ for the Chandra spectrum and at $11.8\sigma$ for the XMM-Newton spectrum. In Figure 3, we compare the "free line" model to the "fixed line' model, including the residuals and the effective area curves. 

\begin{figure*}
\begin{center}
\includegraphics[width=13cm]{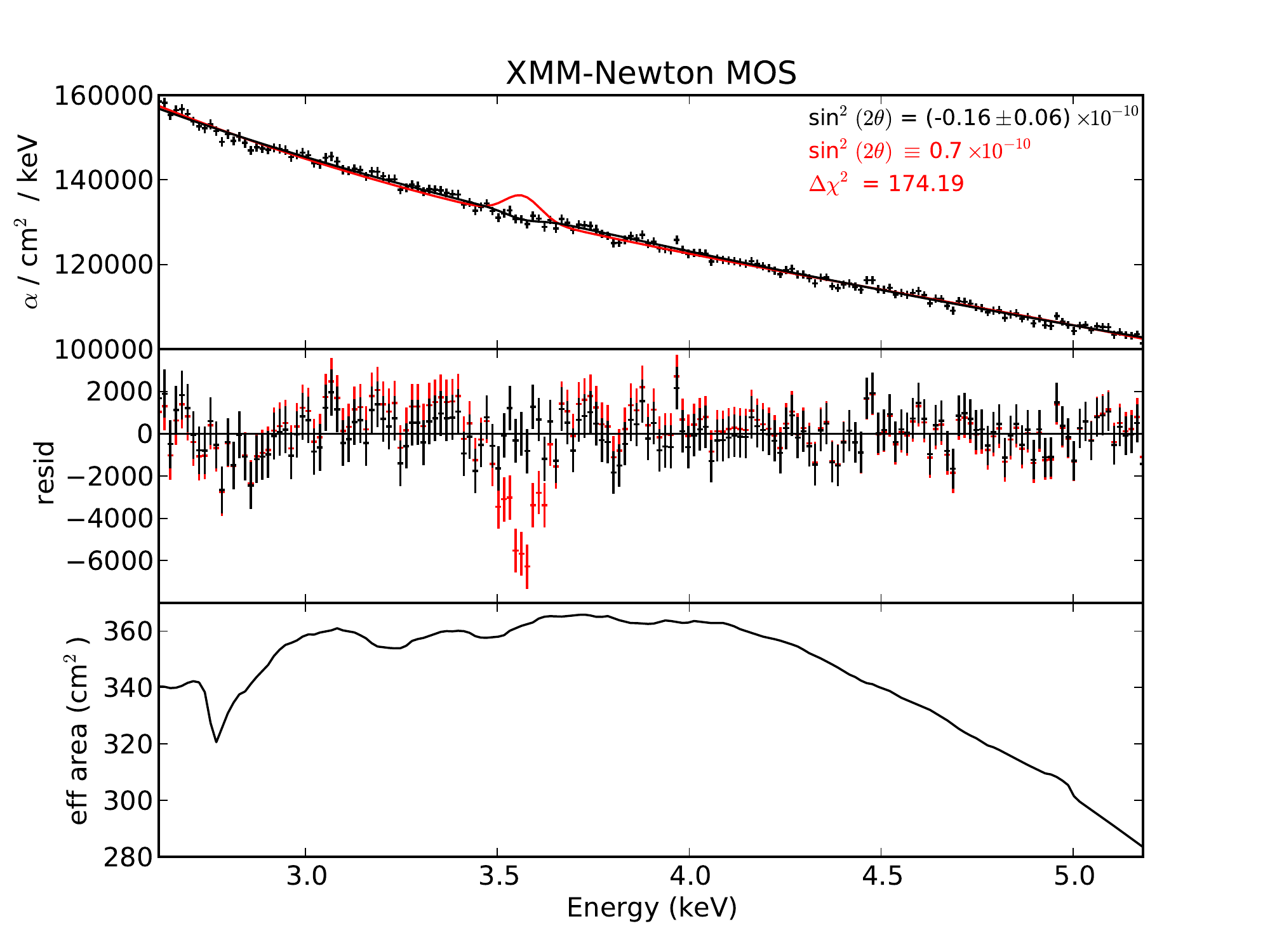}
\includegraphics[width=13cm]{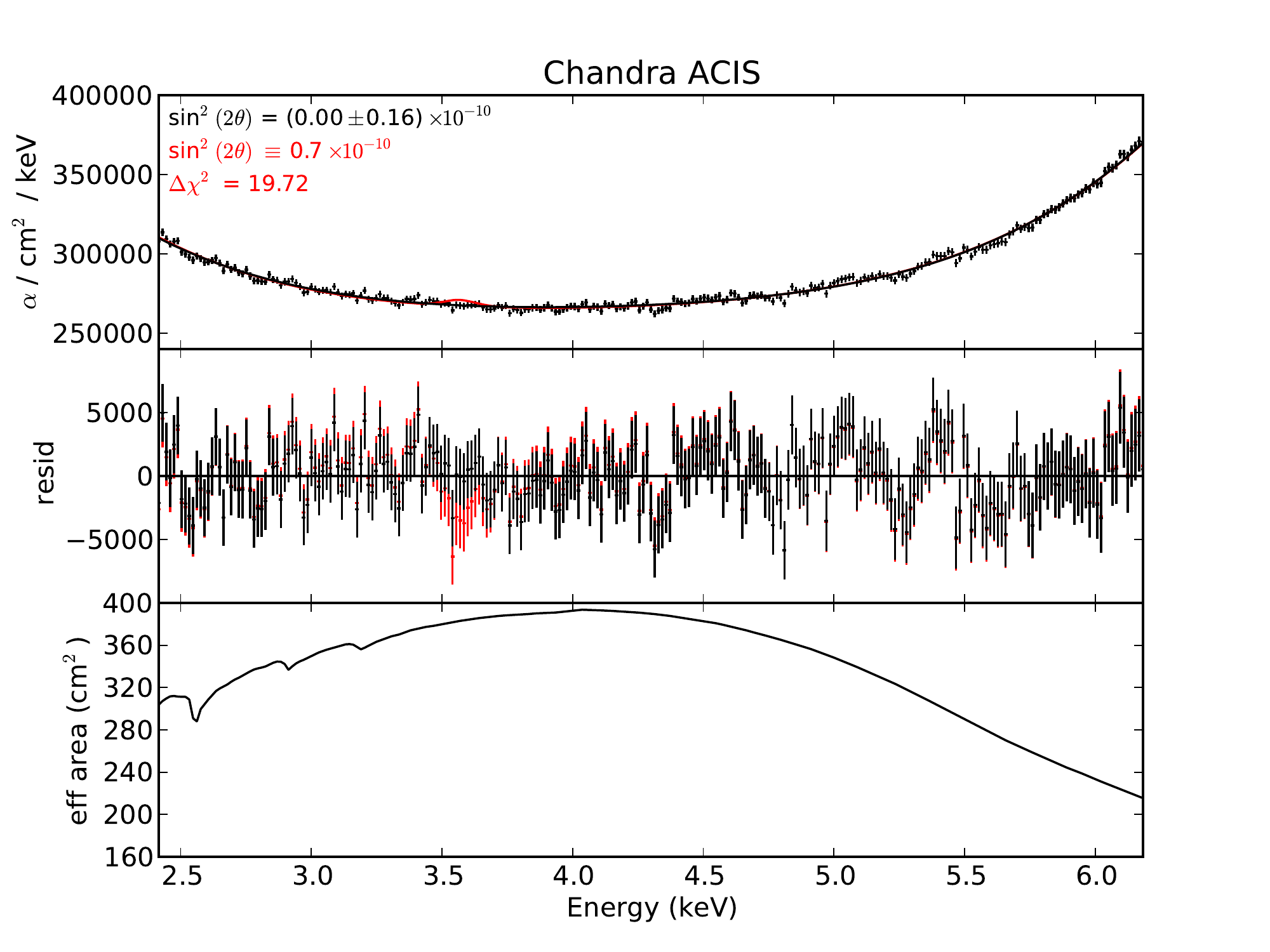}
\end{center}
\caption{Stacked spectra and best-fit spline model (top), residuals (middle), and effective area curves (bottom) of stacked Chandra and XMM-Newton MOS spectra. The black curve indicates a fit with the normalization of the line allowed to float across positive and negative values (the "free line" model); in the red curve the normalization of the  was fixed to the best-fit value from Bu14 (the "fixed line" model). The upper panels also list the mixing angle for each model and the difference in $\chi^2$ between the two models. For both spectra, the best-fit mixing angle is consistent with zero within $3\sigma$. On the other hand, using an F-test, we find the "fixed line" model is ruled out at $11.8\sigma$ for the XMM-Newton spectrum and $4.4\sigma$ for the Chandra spectrum. }
\end{figure*}

We also consider fits where the energy of the line is allowed to vary between 3.50 and 3.60 keV, instead of fixing the energy at 3.57 keV. This accounts for the observed spread in line energies between the various observations of different subsamples in Bu14 and Bo14. For the XMM-Newton spectrum, the best-fit energy for the line is right at 3.50 keV, but the "fixed line" model is still disfavored at $11.0\sigma$. The Chandra spectrum also prefers an energy of 3.50 keV, and in this case the "fixed line" model is only disfavored at $2.7\sigma$. The instrumental line width is comparable to the width of this window, so it is not obvious whether the energy counts as an additional degree of freedom here, but either way the statistical significance remains the same to the first decimal place.

These constraints apply to the assumed mixing angle of $7\times10^{-11}$, which is the best-fit value from Bu14. If we instead consider the lower end of the range measured by Bo14, $2\times10^{-11}$, we find that a line at 3.57 keV with this mixing angle is excluded at 5.1$\sigma$ from the XMM-Newton data.

\section{Analysis}

Finally, we analyze the spectra shown in Figure 3 in order to constrain the possible emission from sterile neutrinos. To do this, we add a \verb"Gaussian" component to our spectral model in similar fashion as Figure 3. The width of this component is set to zero, but we vary the normalization and energy of the line. At each energy, we vary the normalization of the line (allowing for negative normalizations as well) and perform a joint line+spline fit to the data. Using the $\chi^2$ statistic, we find the best-fit normalization for a line if one existed at that energy, as well as the $1\sigma$ and $3\sigma$ allowed intervals. 

We present these results in Figure 4, in a format which allows for consideration of systematic and statistical uncertainties. The $1\sigma$ statistical uncertainties are drawn around the best-fit value for the inferred mixing angle at each energy. The smooth lines denote the $3\sigma$ sensitivity curves, as inferred from the difference between the best-fit value and the $3\sigma$ limits at each energy. The systematic uncertainties can be assessed by comparing the large-scale fluctuations in the best-fit values to these sensitivity curves. Excursions of the best-fit value outside these bounds represent ''detections'' which are formally significant at $\ge 3\sigma$. Such excursions occur at 10.2 keV and 12.2 keV in the stacked Chandra spectrum and at 5.3 and 5.6 keV in the stacked XMM-Newton spectrum. 

However, none of these excursions have the narrow widths that would be expected if they were astrophysical lines. Rather, they appear as broad, large-scale variations and we interpret them as systematic variations in the instrumental background. The lack of coincidence between these excursions in the two spectra supports this interpretation as well, as does the similarity in number and magnitude between positive and negative excursions.  At energies where these variations extend beyond the $3\sigma$ statistical uncertainties, systematic uncertainties are dominant and improved continuum modeling could conceivably improve the sensitivity of our limits. On the other hand, pushing down the statistical uncertainties would require a dataset which is more sensitive to sterile neutrino emission, either due to longer effective integration time or to a more dark-matter-dominated sample. 

As for 7.1 keV neutrinos (corresponding to 3.5-3.6 keV line emission), Figure 4 supports our claims in Figure 3 that we see no evidence of emission lines around this energy. In the Chandra spectrum, there is a weak ($<3\sigma)$ positive residual at about 3.45 keV which is not seen in the XMM-Newton spectrum, but the best-fit mixing angles from both Bo14 and Bu14 are ruled out at $>3\sigma$. In the XMM-Newton spectrum, there is actually a statistically significant negative residual corresponding at neutrinos of mass 7.1 keV, and both the Bo14 and Bu14 detections are well outside our $3\sigma$ sensitivity curves.

\begin{figure*}
\begin{center}
\includegraphics[width=13cm]{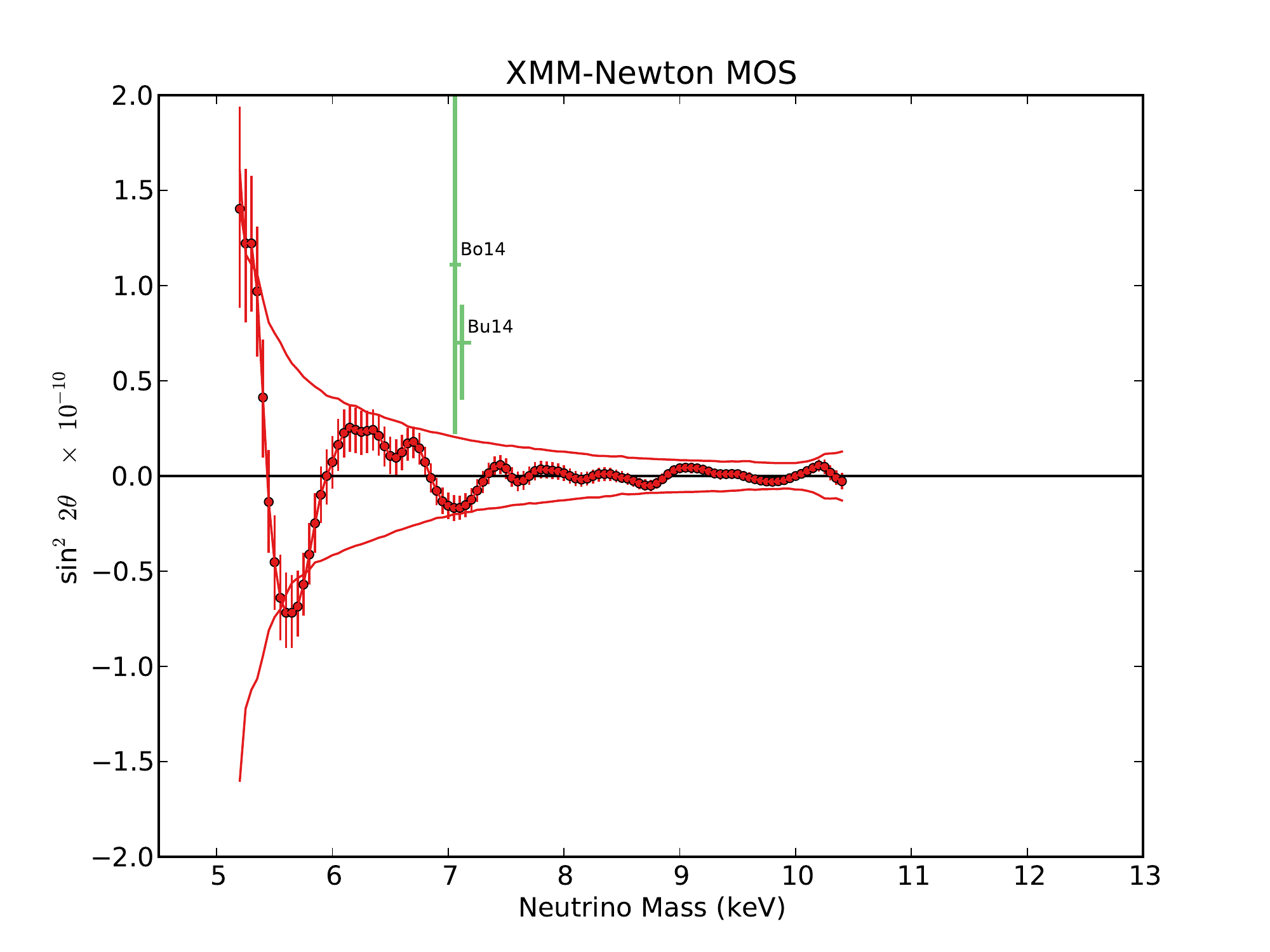}
\includegraphics[width=13cm]{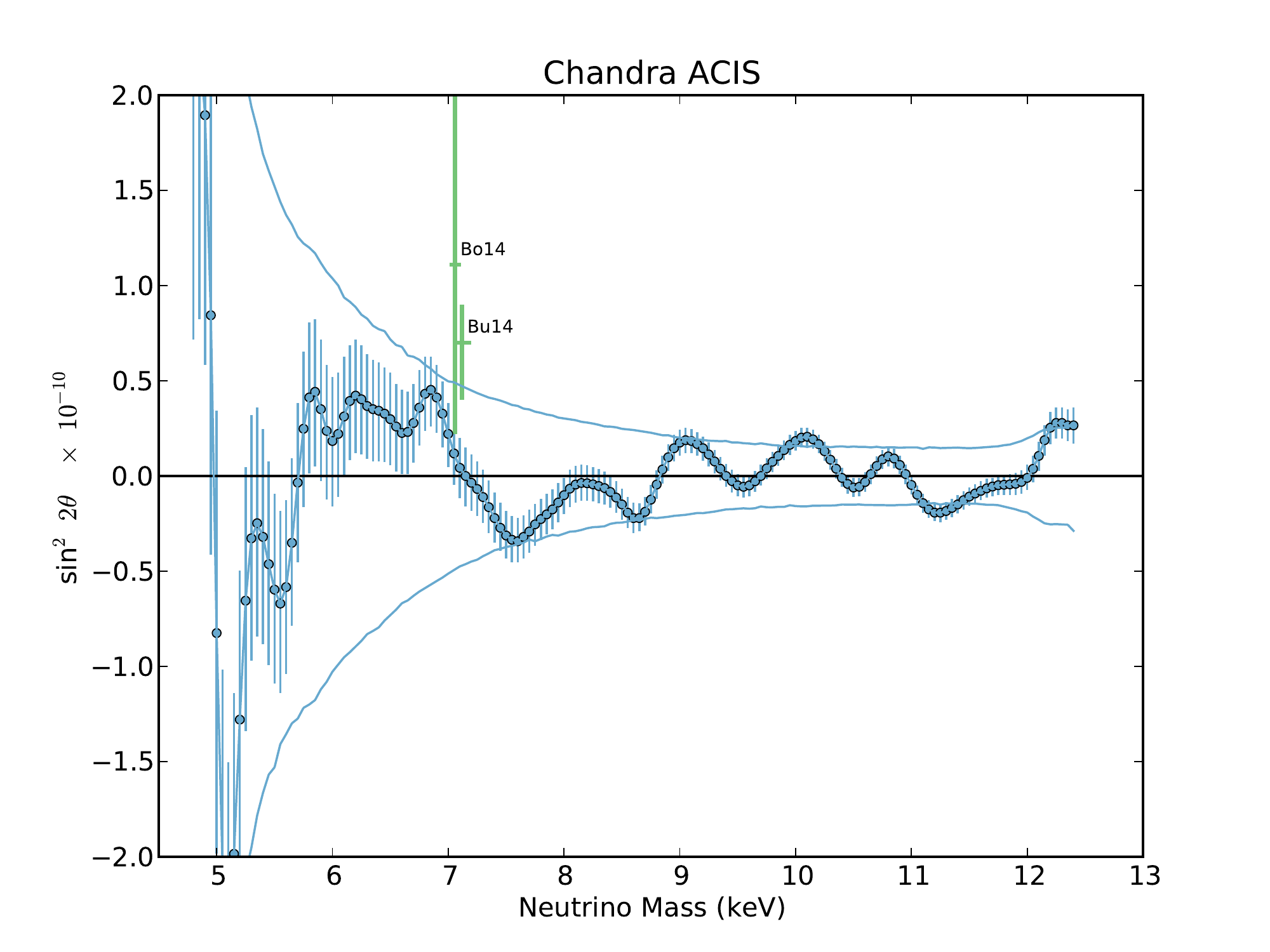}
\end{center}
\caption{Constraints on the mixing angle of sterile neutrinos from the $\nu$MSM model, as derived from our stacked XMM-Newton (top) and Chandra (bottom) spectra. For each potential neutrino mass, we fit a series of Gaussian + spline models to the observed spectrum, with the mean of the Gaussian corresponding to the energy of the photon emitted by decay of sterile neutrinos of the indicated mass. The best-fit normalization of the Gaussian has been converted into a mixing angle and plotted above, as well as the $1\sigma$ confidence interval around the best-fit. The curves are approximate $3\sigma$ sensitivity curves, showing the sensitivity of our experiment as a function of neutrino mass. We interpret excursions outside these curves as systematic errors due to our features in the instrumental background which our spline failed to fit. Better modeling could reduce these uncertainties, but our constraints are already sufficient to rule out the possibility of 7.1 keV neutrinos with the mixing angles suggested by Bu14 and Bo14. }
\end{figure*}

It is important to emphasize that upper limits only conflict with Bu14 and Bo14 if the emission lines they detect are interpreted as sterile neutrino emission. Our nondetections of the same feature in galaxy spectra suggest that their emission lines have an astrophysical origin, possibly some missing piece of intracluster medium astrophysics (e.g. \citealt{Jeltema2014}).

There has also recently been some debate about the possible presence of a line in the centers of the Milky Way and of M31. In the Milky Way, \citet{RiemerSorensen2014} examined 825 ks of Chandra observations of the Galactic center. This study finds no 3.5 keV feature in the combined diffuse Galactic spectrum, and is able to place 95\% upper limits on the mixing angle of sterile neutrinos which are lower than the Bu14 and Bo14 results. Taken at face value, these upper limits are comparable to our limits (lower at some energies, higher at others), although they are significantly more model-dependent. The fields analyzed in this work all fall within 1\% of the virial radius of our Galaxy, which is the region we exclude from our analysis because the results are so dependent on the assumed central logarithmic slope of the dark matter profile. Additionally, the Galactic center is suffused with a very large number of sources of non thermal X-ray emission, so the background at 3.5 keV is quite significant. Riemer-Sorensen therefore uses a line-free APEC model with dozens of Gaussian components to model known emission lines, but this model is still unable to produce statistically acceptable fits to the observed spectra (reduced $\chi^2 = 1.7$ and 5.2 for the two energy bands considered). 

The model-dependence of this result could possibly explain how \citet{Boyarsky2014b} do see a line at 3.53 keV with flux consistent with Bu14 and Bo14 from the Galactic Center. This study also increases the significance of their detection of this line from the center of M31. On the other hand, \citet{Jeltema2014} examine the center of M31 and do not see an emission line. It is outside the scope of this paper to reconcile these various claims, but this disagreement underscores the difficulty of modeling the X-ray background and the dark matter profile in the centers of galaxies, and supports our decision to exclude these regions from our analysis.

\section{Conclusions}

In this work we have presented a comprehensive search for X-ray emission lines from sterile neutrinos. We focused on galaxies and galaxy groups (with $kT \lapprox 1$ keV) in order to avoid contamination from a hot intracluster medium. This sample selection (coupled with point source masking) leads to spectra which are dominated by the instrumental background. We examine 81 objects observed with Chandra and 89 observed with XMM-Newton, yielding total integration times of 15.0 and 14.6 megaseconds, respectively. 

We extract spectra from large annuli around each object, spanning the range (0.01-1.0)$\times R_{\text{vir}}$. We divide each annulus in pixels, and weight each pixel by the expected projected dark matter column density within that pixel. We then stack the spectra from each galaxy, this time weighting the spectra in order to maximize S/N. 

We study the stacked spectra within energy bands which avoid prominent instrumental lines. Using B-splines, we can readily fit the instrumental background and distinguish emission lines from the background. We are able to rule out lines near 3.57 keV with the mixing angle implied by Bu14 at $4.4\sigma$ using our Chandra data and at $11.8\sigma$ using our XMM data.  Allowing the line energy to vary from 3.50-3.60 keV, the Chandra constraint is still $2.7\sigma$ and the XMM-Newton constraint is still $11.0\sigma$. These limits are based on statistical uncertainties, but we show in Appendix B that these results are reasonably robust against systematic errors as well. At 3.57 keV, under the most conservative estimates about systematic uncertainties, we still should have detected the line with the Bu14 mixing angle at $11.8\sigma$ and $4.4\sigma$ respectively.

We extend our search to other energies, and are able to place strong and robust constraints on possible decaying dark matter emission. The limits are shown in Figure 4 for both XMM-Newton and Chandra, and apply to sterile neutrinos with masses ranging from 4.8-12.4 keV. These limits, unlike previous studies, do not depend on assumptions about the X-ray background (since we fit it with B-splines instead of physical models). Moreover, we exclude the centers of the dark matter halos in our sample, so we are also not sensitive to assumptions about the logarithmic slope of the dark matter profiles in this regime. 

During the refereeing process for this paper, a number of studies were released which produced complementary results to our own. In Figure 5, we compare a number of these results, focusing on the studies which produced constraints on the mixing angle as a function of energy instead of just testing one or two specific energies. Each of these broad-band studies produced upper limits. If the Bu14 and Bo14 detections are interpreted as sterile neutrino emission, they are in tension with these upper limits, at various levels of significance. \citet{Malyshev2014} stacked XMM-Newton observations of dwarf spheroidal galaxies, and produced 2$\sigma$ upper limits which are close to our Chandra limits. \citet{Sekiya2015} examined the Suzaku sky background and placed similar upper limits. Their analysis accounted for the look-elsewhere effect (LEE) explicitly as well; we plot both their LEE-corrected and LEE-uncorrected curves for comparison. \citet{RiemerSorensen2014} has already been discussed; this analysis produces the tightest formal constraints but depends sensitively on the Galactic DM profile towards the center of the Galaxy. The upper and lower excluded regions indicate previous constraints as summarized by \citet{Canetti2013}, and the excluded region below 1 keV is an approximate limit based on arguments from the phase-space density of local dwarf spheroidals \citep{Tremaine1979}.  A significant portion of the available parameter space has been ruled out, but the possibility of sterile neutrino dark matter remains.

\begin{figure*}
\begin{center}
\includegraphics[width=17cm]{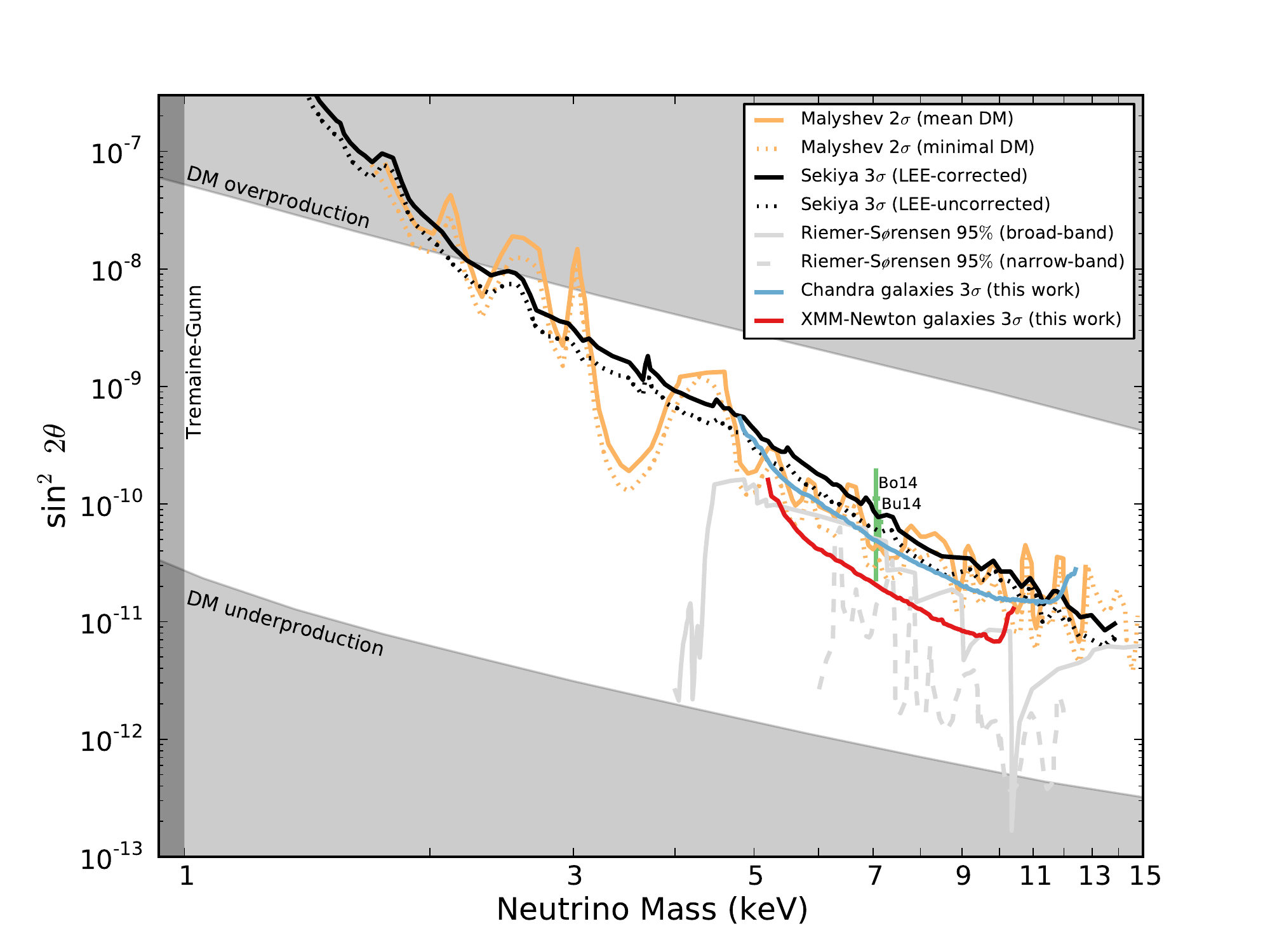}
\end{center}
\vspace{-0.5cm}
\caption{ Upper limits on the mixing angle for sterile neutrinos, from this work (red and blue curves) as well a number of recent studies, indicated in the legend and in the text. If the Bu14 and Bo14 detections are interpreted as sterile neutrino emission, they are in tension with these upper limits, at various levels of significance. A few constraints from the literature are also displayed with the gray shaded regions, as discussed in the text. A significant portion of the available parameter space has been ruled out, but the possibility of sterile neutrino dark matter remains. }
\end{figure*}

\section{Acknowledgements}
\vspace{-0.2 cm}

The authors would like to thank D. Prokhorov and R. Sunyaev for helpful discussions and insight during the writing of this manuscript. This research has made use of the NASA/IPAC Extragalactic Database (NED) which is operated by the Jet Propulsion Laboratory, California Institute of Technology, under contract with the National Aeronautics and Space Administration. This research has made use of NASA's Astrophysics Data System. We acknowledge the usage of the HyperLeda database (http://leda.univ-lyon1.fr). 

\bibliographystyle{mn2e}

\appendix
\counterwithin{figure}{section}

\section{Varying Assumed Scaling Relations}

For each galaxy, we assume a model dark matter profile which is used to weight the signal from each pixel. We assume the halo obeys an NFW profile (excluding the central 1\% of the virial radius where the shape is more uncertain), so our model has just two free parameters: the total mass of the halo, and the concentration parameter. As discussed in sections 2.1.3 and 2.2, we set these parameters using scaling relations. For the total mass of the halo, we use the abundance matching relation of \citet{Moster2010} to scale from the inferred stellar mass to the inferred halo mass, and for the concentration we use the $M-c$ relation of \citet{Prada2012} to scale from the halo mass to the concentration parameter (assuming $z=0$). 

In this Appendix, we test the dependence of our results on these assumptions, by adopting different scaling relations and repeating our analysis. For the abundance matching relation, we examine the \citet{Behroozi2010} relation (their fiducial relation, allowing for systematic offsets in the inferred stellar mass). For the mass-concentration relation, we use the Halo Evolution Web Calculator \citep{Zhao2009} assuming a BBKS1986 \citep{Bardeen1986} power spectrum and WMAP7 cosmological parameters \citep{Komatsu2011} to generate a list of masses and concentrations for $z=0$, and we linearly interpolate between these values to derive concentrations for each galaxy we examine.

The results can be seen in Figs. A1 and A2. The sizes of the uncertainties are affected somewhat, but our overall conclusions are entirely unchanged. 

\begin{figure} 
\begin{center}
\subfigure{\includegraphics[width=6.8cm]{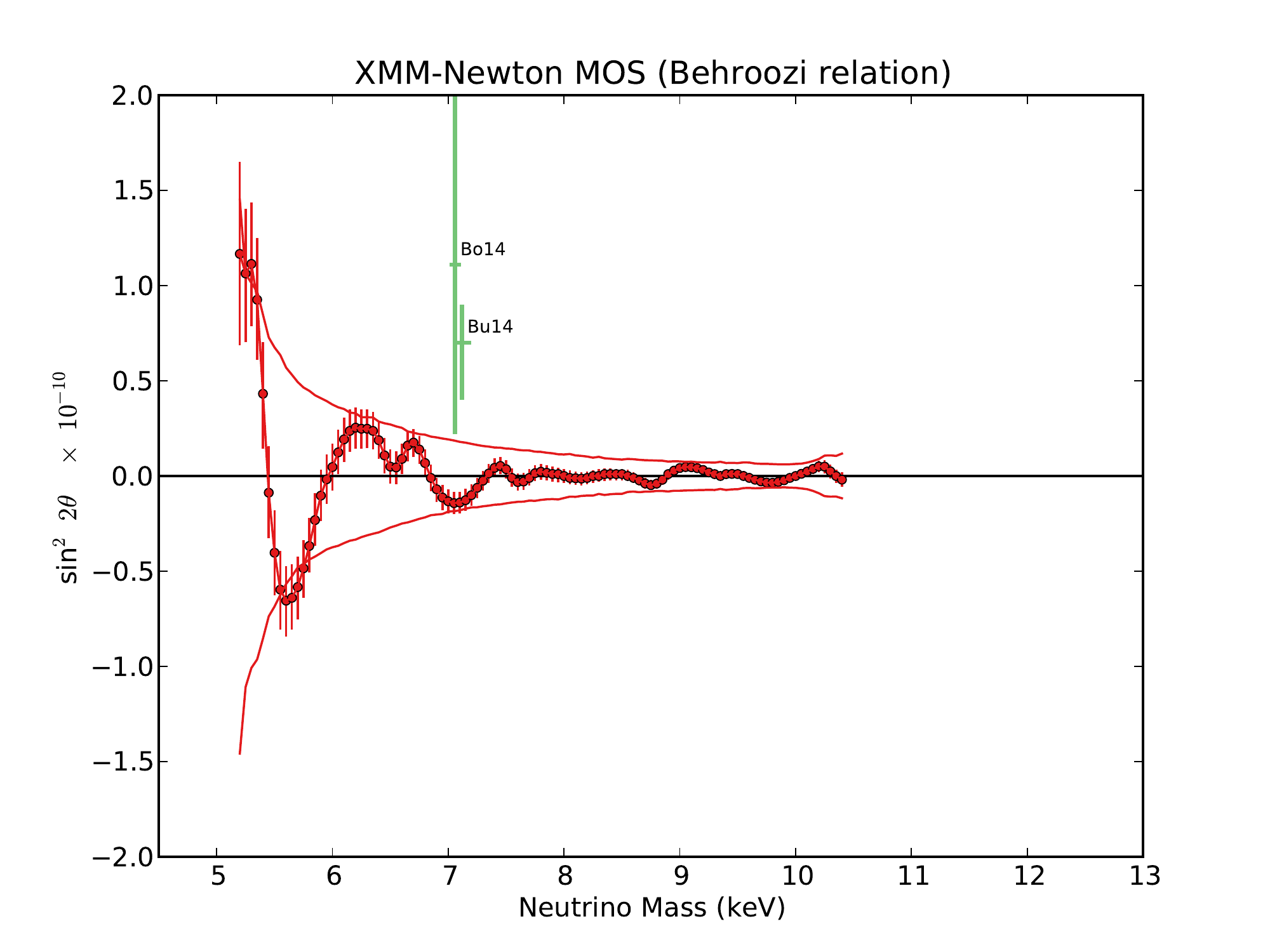}}
\subfigure{\includegraphics[width=6.8cm]{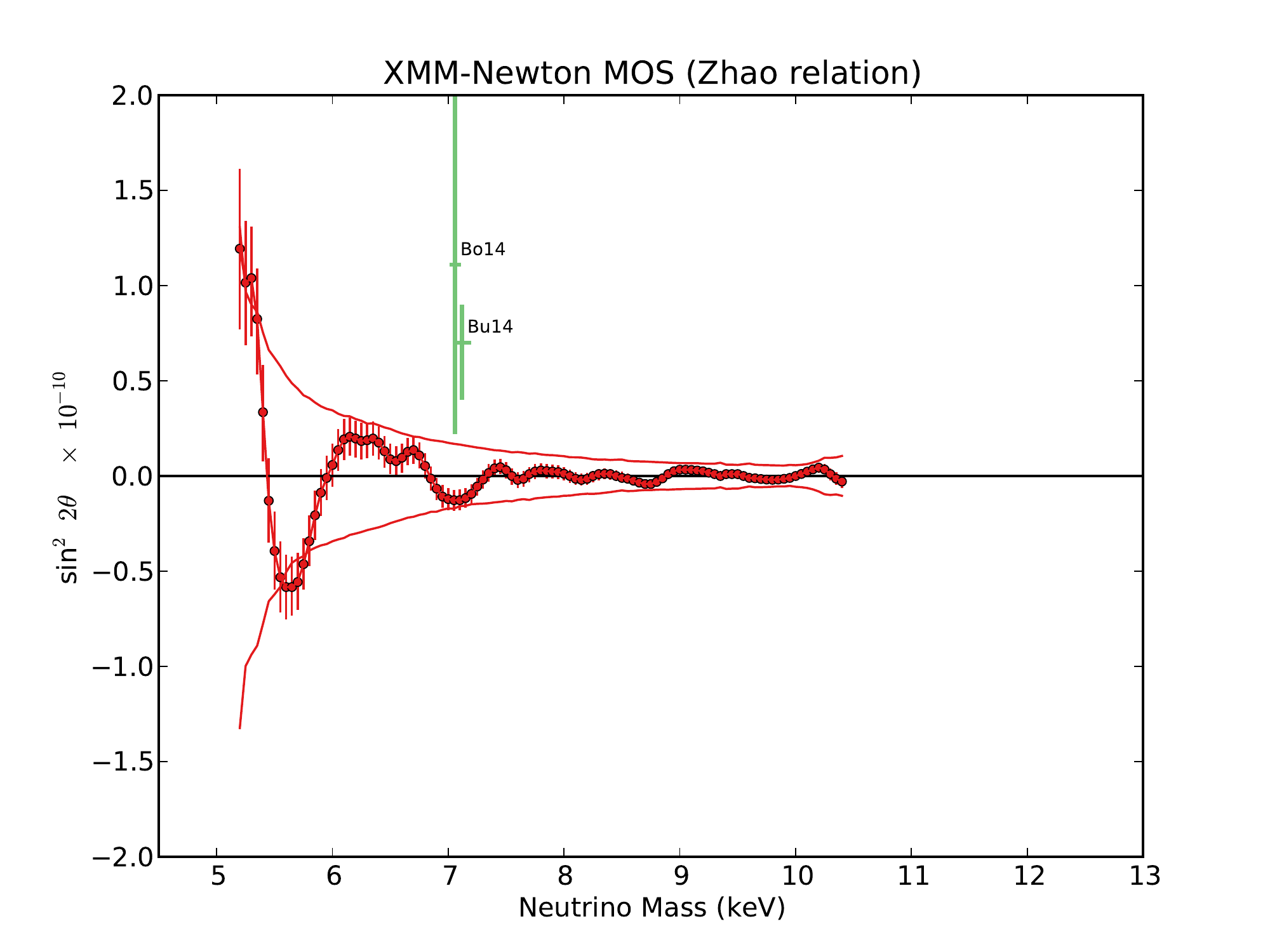}}
\subfigure{\includegraphics[width=6.8cm]{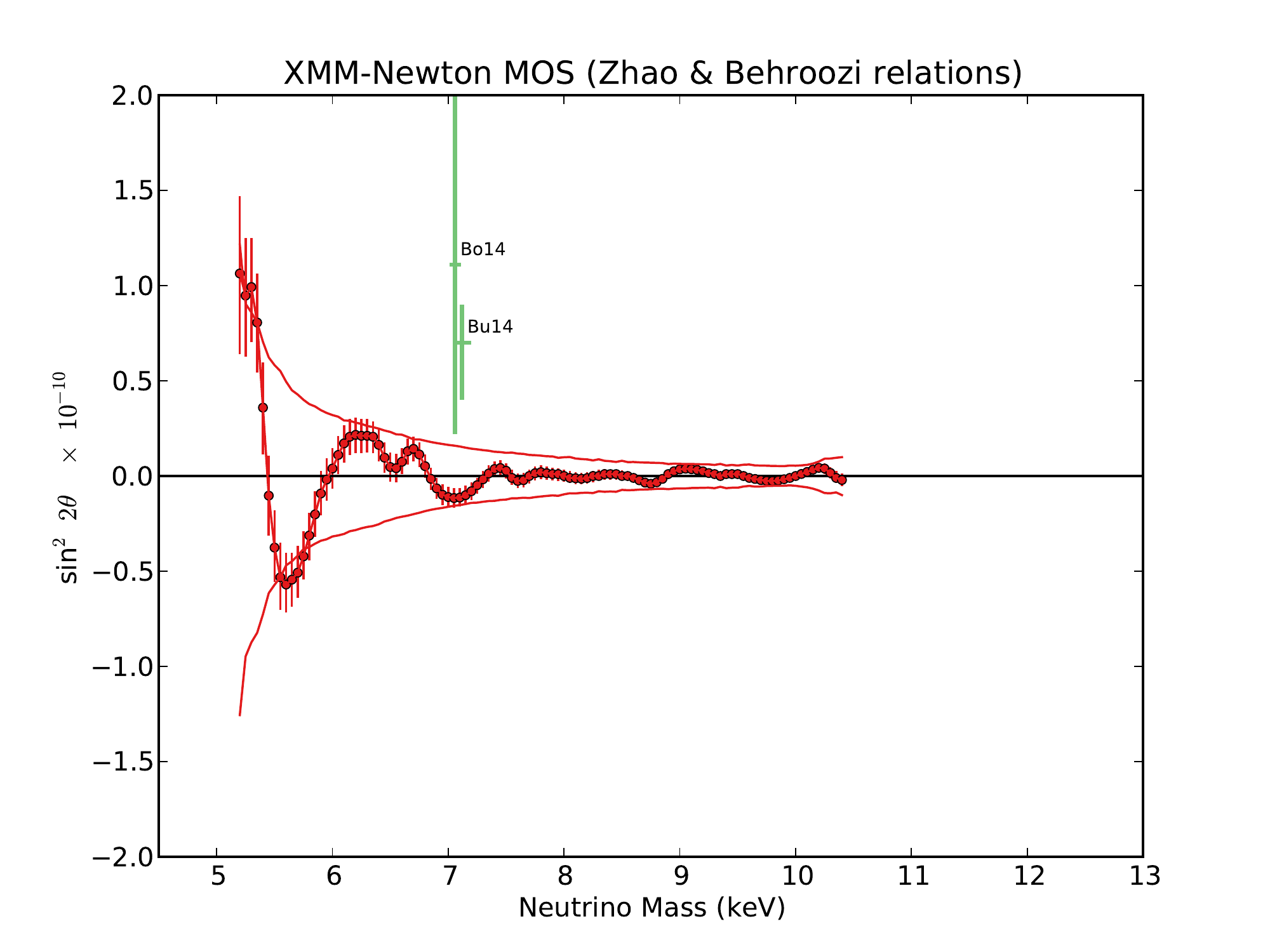}}
\end{center}
\vspace{-0.5cm}
\caption{Same as Figure 4, but using alternate scaling relations to fix the assumed halo mass and concentration for each galaxy. }
\end{figure}

\begin{figure} 
\begin{center}
\subfigure{\includegraphics[width=6.8cm]{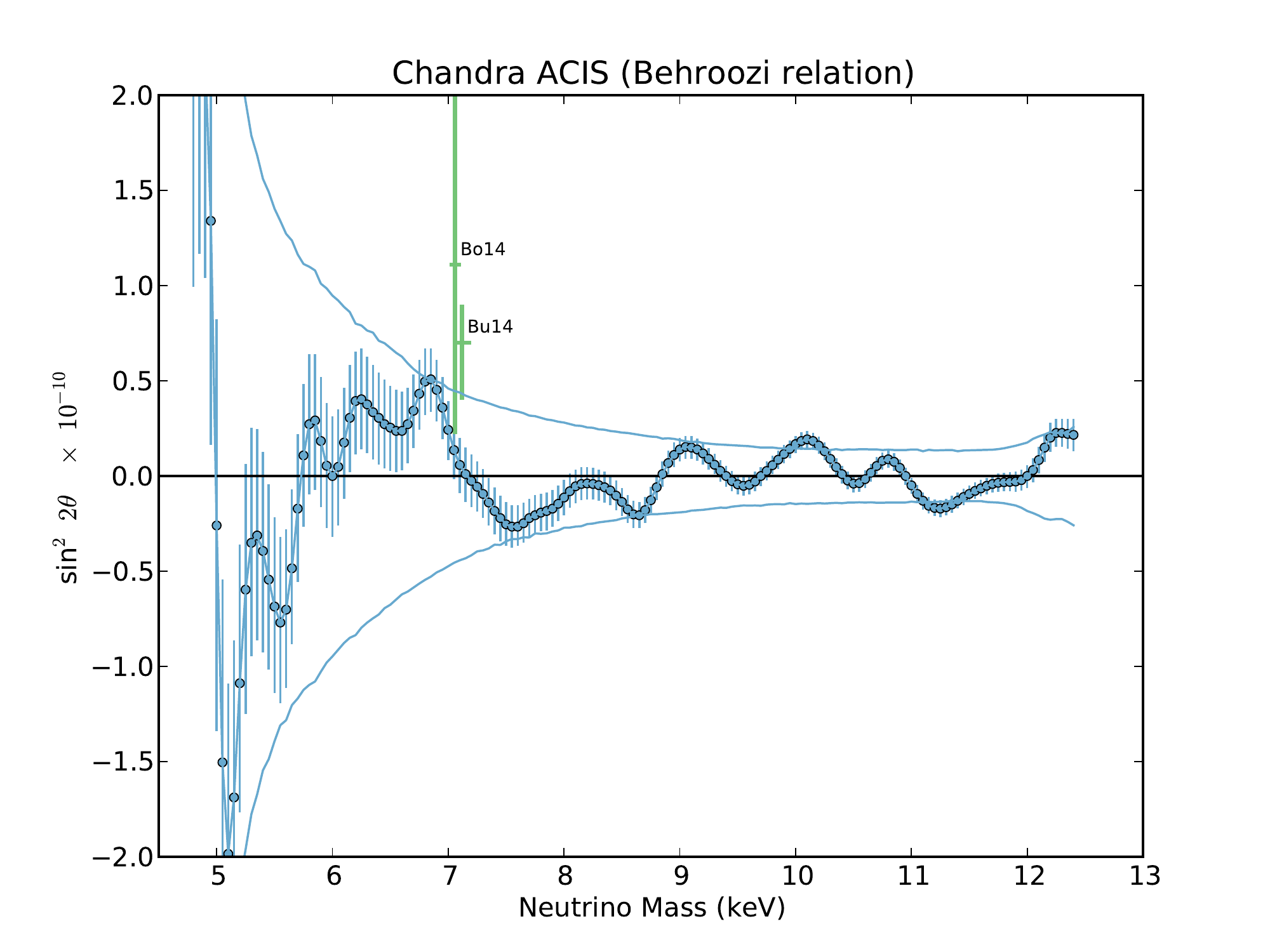}}
\subfigure{\includegraphics[width=6.8cm]{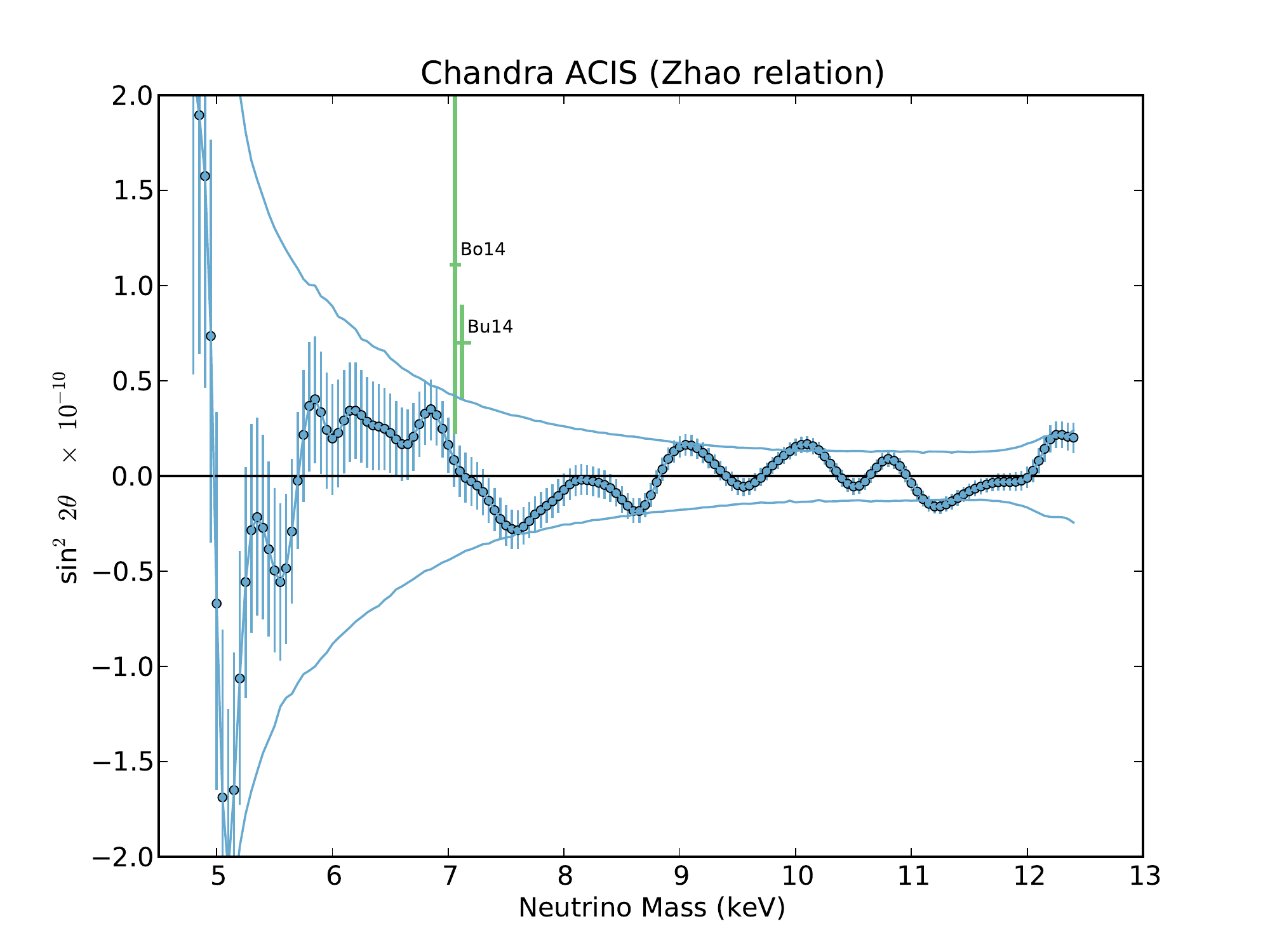}}
\subfigure{\includegraphics[width=6.8cm]{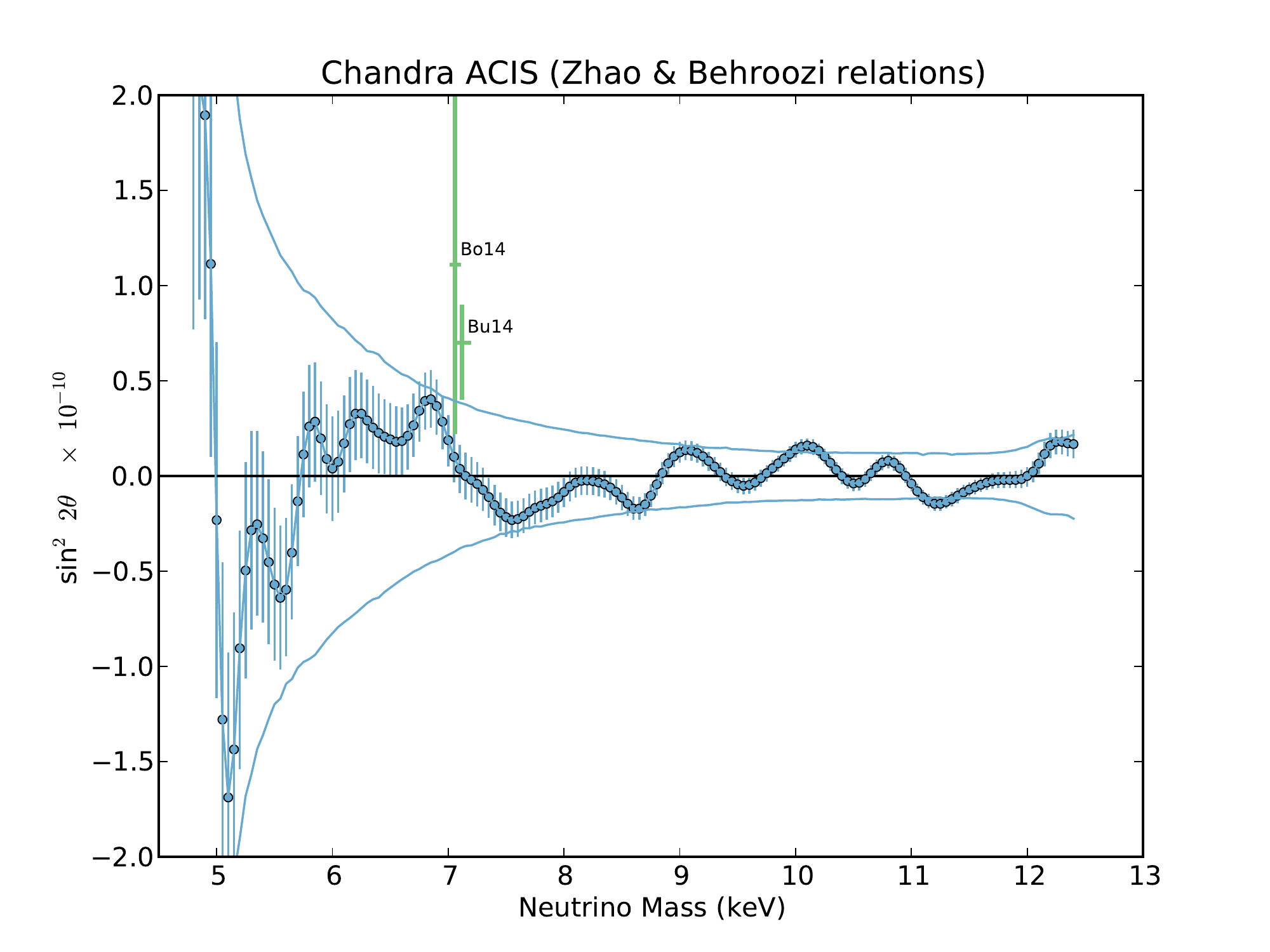}}
\end{center}
\vspace{-0.5cm}
\caption{Same as Figure 4, but using alternate scaling relations to fix the assumed halo mass and concentration for each galaxy. }
\end{figure}

We can also analytically explore the effect of errors in the estimated dark matter content for our sample galaxies. Let us assume a model $M_i$ for the dark matter column density profile, while the true dark matter column density profile is instead $M_i'$. Assume the image flux is uniform (since our images are dominated by instrumental backgrounds) with flux per pixel $I_i$. Then, from eqs. (7)-(9), the estimated weighted value of $\alpha$ is

\begin{equation}
\alpha  = \frac{\sum_i{(I_i / M_i)  \frac{1}{\sigma^2_{\alpha_i}}}}{\sum_i{\frac{1}{\sigma^2_{\alpha_i}}}}
\end{equation}

Substituting $\alpha' M_i'$ for $I_i$, where $\alpha'$ is the true value of the estimator (in contrast to the measured value $\alpha$), and noting that $\sigma_{\alpha_i} \propto 1/M_i$ (the constant of proportionality is $\sigma_{I_i}$, which is assumed to be the same in every pixel and cancels out of the equation):

\begin{equation}
\alpha  = \alpha' \frac{\sum_i{\frac{M_i'}{M_i} M_i^2}}{\sum_i{M_i^2}} = \alpha'  \frac{\sum_i{M_i'}{M_i}}{\sum_i{M_i^2}} 
\end{equation}

Using this relation, we can compute analytically the bias in the inferred value of $\alpha$ due to errors in the model for the dark matter profile. We find that this bias is small, especially if we integrate all the way out to the estimated virial radius. Due to the $M/R^2$ prefactor in eq. (10), the column density only scales as $M^{1/3}$, so a $30\%$ bias in the total dark matter mass of a galaxy leads only to a $10\%$ bias in the estimator $\alpha$ for that galaxy. A change in NFW concentration parameter has even less of an effect, since the change in mass at small radii is compensated by the change in mass at large radii. Integrating out to the virial radius, changing $c$ from 8 to 6 only lowers $\alpha$ by $0.5\%$ (and changing it from 8 to 10 increases $\alpha$ by $0.5\%$). If we only integrate out to half the virial radius, the effect is larger (rising to the $3\%$ level), but still not very significant. 

It is also important to emphasize that these errors are measured for individual galaxies. As long as the abundance matching and $M$-$c$ relations are not biased, stacking will reduce these errors in aggregate, since the positive outliers will cancel with the negative outliers. Thus we expect even better accuracy for the total sample then the above estimates for individual objects.

\section{Verifying Flux Recovery for our Spline Fitting}

Since our analysis uses a spline to fit the spectrum, it is important to verify that the spline is able to distinguish a spectral line from a variation in the continuum. To verify this, we added spectral lines (with noise, folded through the instrumental response) to both the Chandra and XMM stacked spectra. We injected these lines at a number of different energies, and at each energy we repeated the analysis for lines corresponding to three different mixing angles ($7\times10^{-12}$, $7\times 10^{-11}$, $7\times10^{-10}$).

For each injection, we fit a Gaussian + spline model to the spectrum, with the Gaussian centered at the energy of the injected line. We compute the best-fit mixing angle at that location, and compare it to the best-fit mixing angle in the original spectrum (without any injected lines). The difference between these two mixing angles is plotted in Figure B1. We repeat this procedure for 100 realizations, taking the central 68\% of the resulting values as the $1\sigma$ confidence interval around the median value. 

These simulations show that there is a systematic effect where the large-scale systematic fluctuations in the spectrum (see Figure 4) slightly bleed into the inferred mixing angles, biasing them upwards or downwards in the same direction as the larger-scale fluctuations. This effect is very small, however. At the Bu14 value of $7\times10^{-11}$, it is comparable to the $1\sigma$ dispersion in the recovered mixing angle for the Chandra spectrum, and even smaller for the XMM spectrum. Even at injected mixing angles $10\times$ smaller, the XMM data approximately obey Gaussian statistics, with 10/16 (63\%) of the simulations overlapping with the true injected mixing angle within the $1\sigma$ dispersion. 

We emphasize that the nature of the systematics here is linear: the spline is simply missing small-scale residuals in the spectrum, so inferred mixing angles at the locations of these residuals are moved upwards or downwards exactly by the size of the residual. For example, for a neutrino mass of 7.14 keV, there is a negative residual at 3.57 keV in the XMM-Newton spectrum, corresponding to a mixing angle of $-1.6\pm0.6\times 10^{-11}$ (see Figure 3). This is itself evidence against the presence of an emission line at this energy, but we can neglect this for the moment to derive the most conservative estimate of the systematic error introduced by this negative residual. For the XMM-Newton spectrum at 3.57 keV, this translates to a reduction in the constraint by a factor of $7^2 / (7+1.6)^2 \approx 0.66$, so the tension with the Bu14 best-fit mixing angle decreases from 11.8$\sigma$ to 7.8$\sigma$ under this conservative estimate of the systematics. The Chandra spectrum has no residuals at 3.57 keV, so this systematic has no effect here.

\begin{figure}
\begin{center}
\subfigure[Chandra]{\includegraphics[width=8cm]{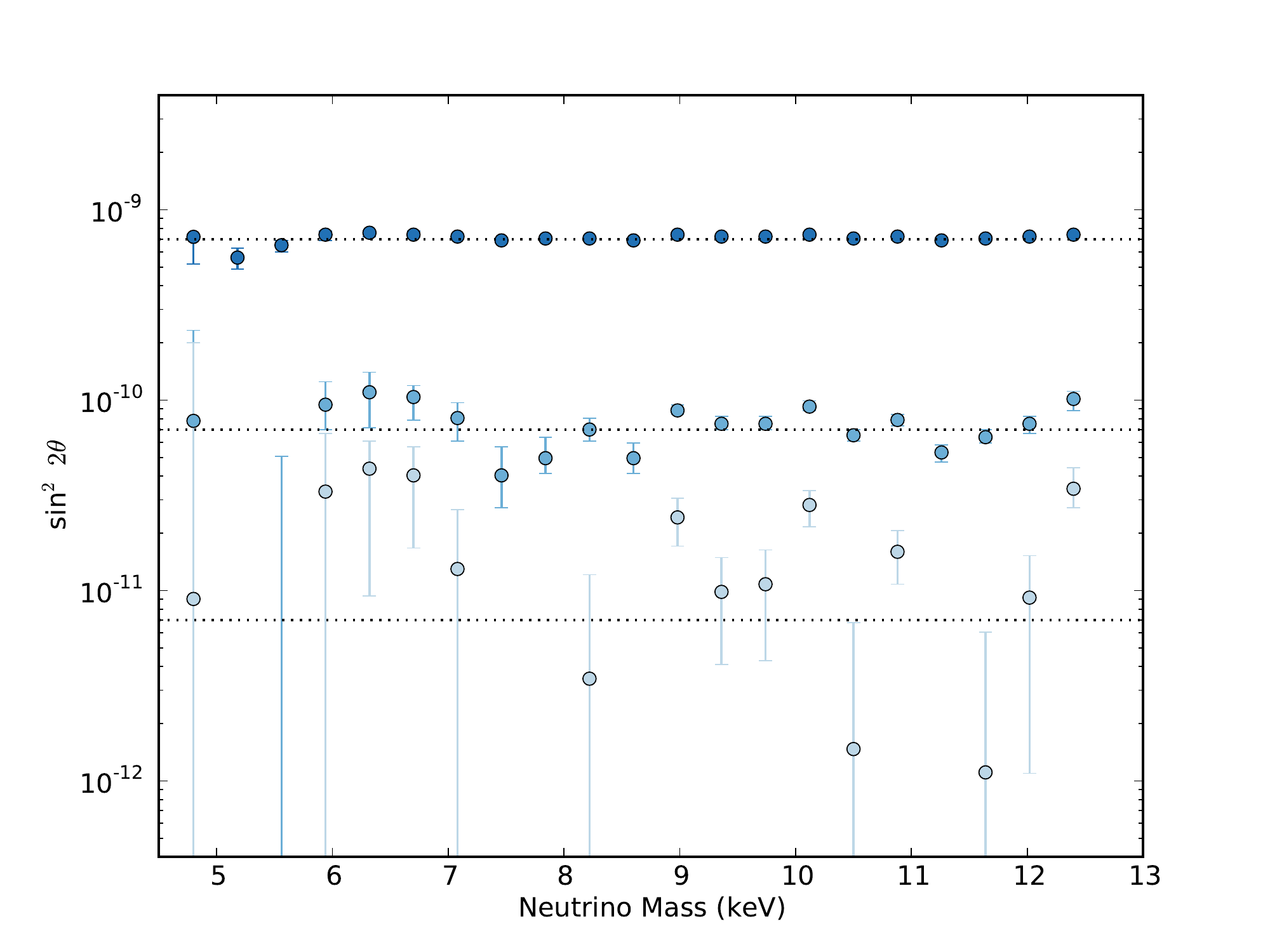}}
\subfigure[XMM]{\includegraphics[width=8cm]{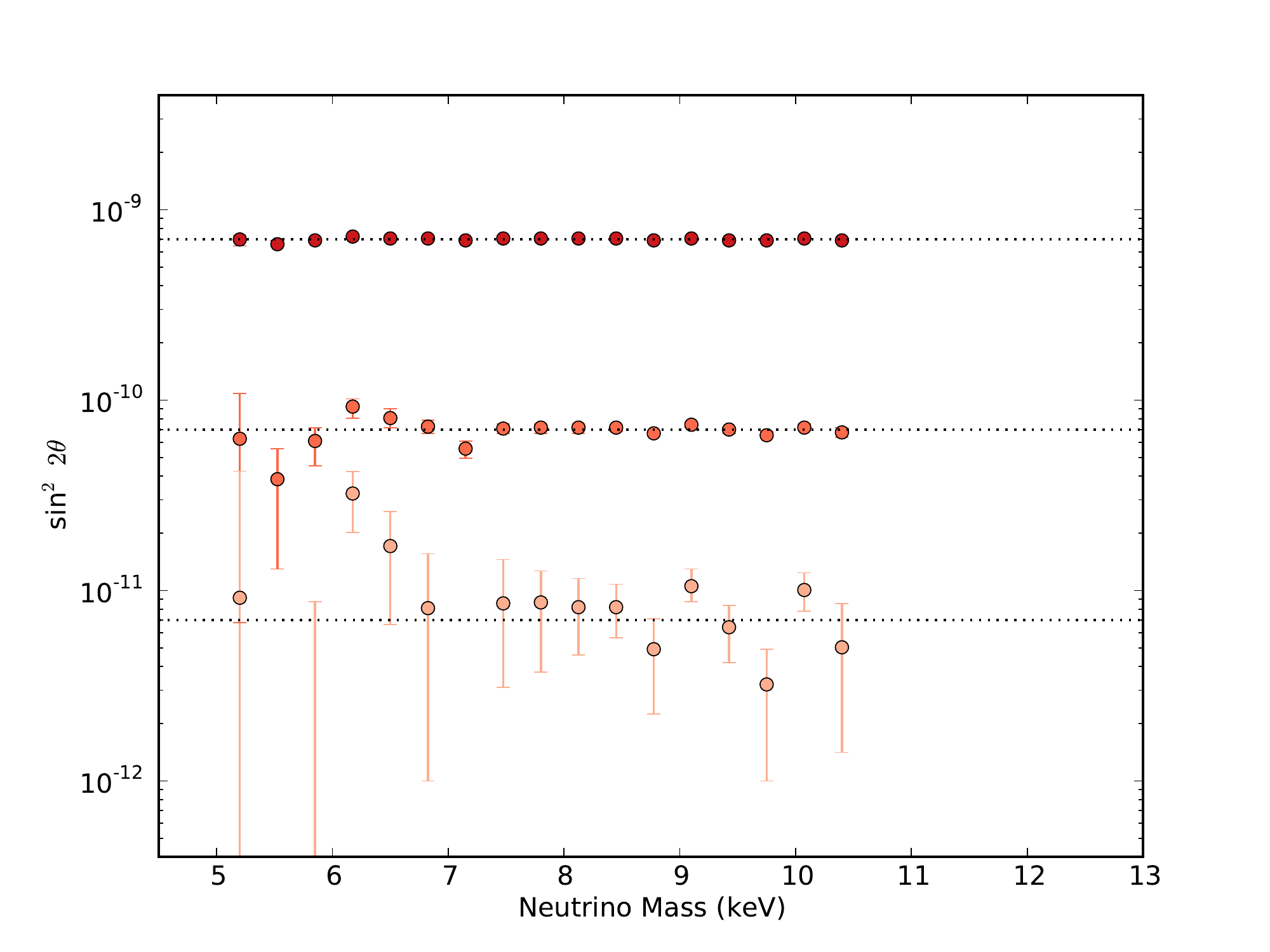}}
\end{center}
\vspace{-0.5cm}
\caption{Difference in recovered mixing angle between original spectrum and spectrum with a simulated emission line injected. Each point corresponds to a different simulated spectrum, with a line injected at the indicated energy. The point shows the difference in the recovered mixing angle, and the lines show the input mixing angle used to generate the simulated lines. Each injection was repeated 100 times, and the error bars show the central 68\% of the resulting values, interpreted as the $1\sigma$ confidence interval around the median value.  }
\end{figure}

Another somewhat related concern is the possibility that the large-scale residuals visible in Figure 4 are not instrumental features, but are instead related to astrophysical signals (most likely unresolved X-ray binaries or the cosmic X-ray background) folded through the instrumental response. If this is the case, it would introduce another source of systematic uncertainty, which would weaken somewhat the degree to which we can rule out the presence of line emission. Here we perform an additional test to check the importance of astrophysical emission to the residuals in Figure 4.

For each telescope, we examine a few galaxies with the best expected S/N ratio. We generate the weighted stacked signal for just these galaxies exactly as above, and for comparison we also compute the weighted stacked spectrum from alternative images without the bright point sources masked (although we still mask the central region exactly as before). We then repeat the same analysis as Section 5 on these pairs of spectra, fitting a spline to the spectrum and constraining the mixing angle corresponding to potential lines at energies across the spectrum. The results are shown in Figure B.2. 

The residuals are slightly larger in the spectrum produced from the unmasked image relative to the masked image, suggesting that masking the bright point sources does indeed improve our sensitivity. The question is whether this improvement could be extended with a deeper observation that allows for masking out more of the astrophysical emission from the image. To quantify this, we estimate the fraction of the total astrophysical emission we have masked out in these observations. 

For the five Chandra observations, the mean stacked exposure times within a few arcminutes of the centers of the galaxies are 300 ks, 500 ks, 900 ks, 800 ks, and 800 ks for NGC 3379, NGC 4278, NGC 3115, M33, and NGC 5457 respectively. We assume a mean CXB surface brightness of $2.6\times10^{-11}$ erg s$^{-1}$ cm$^{-2}$ deg$^{-2}$ (in the 0.5-7.0 keV band, estimated by adding up the background components in \citet{Hickox2006}; note that this band includes some Galactic halo emission as well, which can be treated the same as the CXB for this analysis). For a fiducial circular mask of radius 1 arcsecond (i.e. moderately close to the aimpoint), we can therefore expect 0.15-0.45 counts from the CXB. Given the false source detection probability of $1\times10^{-6}$ used for the \verb"wavdetect" point source detection algorithm, this translates to a requirement of 4-6 counts (0.5-7.0 keV band) within the region in order for a point source to be detected and masked, or an effective flux limit of about $1-2\times10^{-16} $ erg s$^{-1}$ cm$^{-2}$ in the same band.

The most distant galaxy of these five is NGC 4278 ($d = 15.8$ Mpc); at this distance the conservative flux limit of $2\times10^{-16}$ erg s$^{-1}$ cm$^{-2}$ corresponds to $L_X = 5\times10^{36}$ erg s$^{-1}$. Examining the cumulative X-ray luminosity function of low-mass X-ray binaries, cataclysmic variables, and coronally active binaries \citep{Sazonov2006}, we can see that a complete survey of point sources down to this limit will resolve more than 90\% of the total luminosity in these sorts of point sources. Thus we have masked almost all the astrophysical emission associated with these five galaxies. For the CXB, our completeness is not quite as good, but from the $N$-$S$ relation in \citet{Lehmer2012} we estimate that a flux limit of $2\times10^{-16}$ erg s$^{-1}$ cm$^{-2}$ resolves $\approx 2/3$ of the CXB. Combining these completenesses, we see that most of the astrophysical emission is expected to be masked, so the fact that the residuals are not significantly changed between the masked and unmasked images shows that the residuals are not primarily caused by astrophysical emission. They are therefore intrinsic to the instrumental background, as we surmised, and we do not expect that they can be improved significantly with further improvements in masking or modeling the astrophysical backgrounds.

With XMM-Newton, the five galaxies with the best expected S/N ratio span a much wider range in distance and completeness, but we can get a good picture of the residuals just by examining M31, since it contributes the most to the overall signal out of any galaxy we examined. Within the central few arcminutes, the mean stacked exposure time is about 700 ks for this galaxy. The psf is significantly larger than Chandra ACIS, however, so for a fiducial circular mask of radius 20 arcsec, we can expect about 200 counts from the CXB. This translates to a requirement of 275 counts for a detection of a point source, or an effective flux limit of $4\times10^{-15}$ erg s$^{-1}$ cm $^{-2}$. At the distance to M31 (which we take to be 790 kpc), the flux limit corresponds to an XRB with $L_X = 3\times10^{35}$ erg s$^{-1}$. We can therefore expect to mask essentially all of the XRB flux. The CXB completeness is not as good as with Chandra, but M31 is so close that the XRB flux should dominate over the CXB flux, so we can expect to be masking the majority of the astrophysical emission in the M31 field. Just as with Chandra, we see that the residuals and the expected sensitivity are slightly improved with this masking, but not nearly as much as one would expect if they were dominated by astrophysical emission.

\begin{figure}
\begin{center}
\subfigure[Chandra]{\includegraphics[width=8cm]{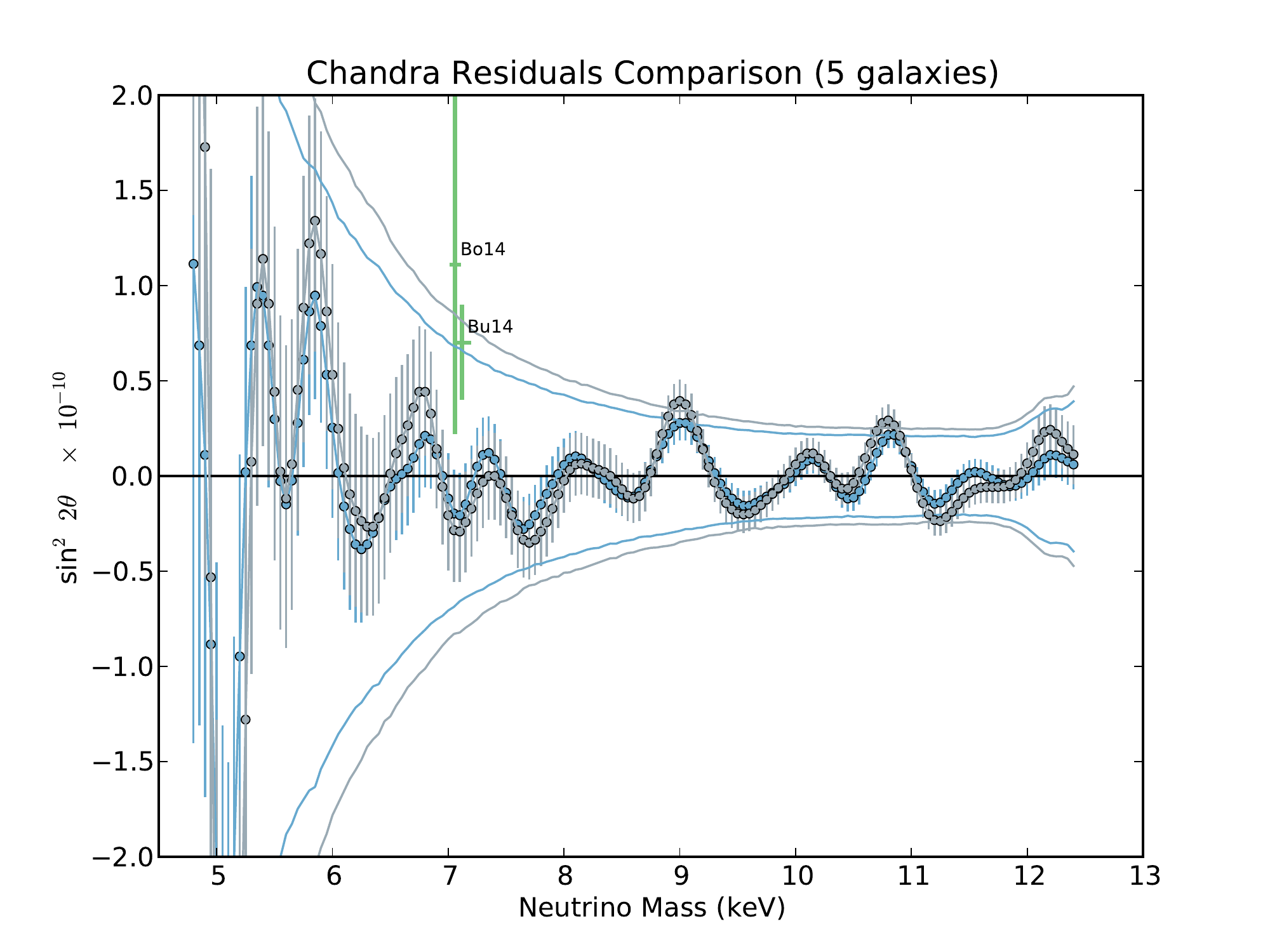}}
\subfigure[XMM]{\includegraphics[width=8cm]{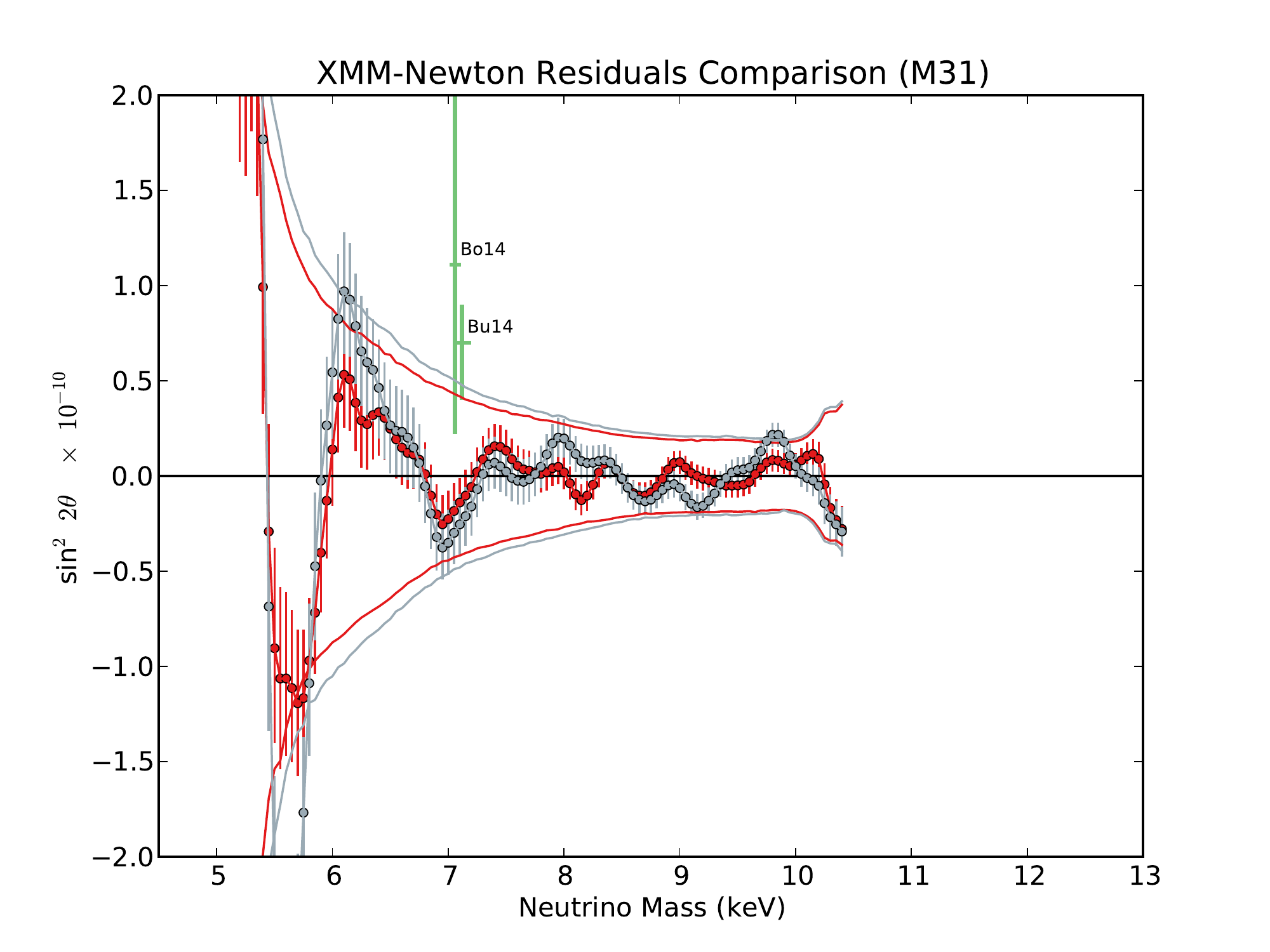}}
\end{center}
\vspace{-0.5cm}
\caption{Same as Figure 4, but using only NGC 3379, NGC 4278, NGC 3115, M33, and NGC 5457 for Chandra (left) and only M31 for XMM-Newton (right). In both plots, the colored line shows the residuals with the same masks used as the fiducial analysis, and the gray line shows the residuals if no point source masking is applied. In both cases, we expect that the masks are capturing most of the astrophysical emission in the field (X-ray binaries and the CXB). Thus the residuals improve somewhat due to the masking, but most of the remaining residuals therefore cannot be attributed to astrophysical sources and are instead likely due to the instrumental background.}

\end{figure}

\section{List of Stacked Galaxies}
\newpage
\begin{deluxetable}{lrrcccccll}
\tabletypesize{\tiny}
\tablecolumns{10}
\tablewidth{0pt}
\tablecaption{ Stacked Galaxies }
\tablehead{
\colhead{name} & \colhead{ra} & \colhead{dec} & \colhead{$t_{\text{CXO}}$}& \colhead{$t_{\text{XMM}}$} & \colhead{d}  & \colhead{log $M_{\text{halo}}$} & \colhead{$R_{\text{vir}}$}  & \colhead{ObsID} & \colhead{ObsID}  \\
\colhead{} & \colhead{} & \colhead{} & \colhead{(ks)}&  \colhead{(ks)}&  \colhead{(Mpc)}  & \colhead{$(M_{\odot})$} & \colhead{(kpc)} & \colhead{(CXO)} & \colhead{(XMM)} }
 \startdata
NGC45     &  3.516625  & -23.182083  & 65.1    &&     9.30 & 11.25 &   110 &  4690,6184,6185&\\
\hline
NGC55    &   3.723333   & -39.196639 & 68.9 &             127.4  &     1.95  & 11.15  &  102& 2255,4744 &  0655050101\\
\hline
M31    &  10.684793 &   41.269065    & 118.1 &        1021.3 &     0.79   & 11.94 &    188   &307,308,309,310,311, &
0109270101,0109270501,\\
&&&&&&&& 312,1575,1577,1581, & 0112570101,0112570401,\\
&&&&&&&&1582,1583,1585,1854, & 0112570501,0112570601,\\
&&&&&&&& 2895,2896,2897,2898,& 0202230201,0202230301,\\
&&&&&&&& 4360 & 0202230401,0202230501,\\
&&&&&&&&&0405320501,0405320601,\\
&&&&&&&&&0405320701,0405320801,\\
&&&&&&&&&0405320901,0505720201,\\
&&&&&&&&&0505720301,0505720401,\\
&&&&&&&&&0505720501,0505720601,\\
&&&&&&&&&0551690201,0551690301,\\
&&&&&&&&&0551690401,0551690501,\\
&&&&&&&&&0551690601,0560180101,\\
&&&&&&&&&0600660201,0600660301,\\
&&&&&&&&&0600660401,0600660501,\\
&&&&&&&&&0600660601,0650560201,\\
&&&&&&&&&0650560301,0650560401,\\
&&&&&&&&&0650560501,0650560601,\\
&&&&&&&&&0674210201,0674210301,\\
&&&&&&&&&0674210401,0674210501,\\
&&&&&&&&&0674210601,0700380501,\\
&&&&&&&&&0700380601,0727960401\\
\hline
NGC247    &  11.785625  & -20.760389     &&       51.7     &   3.59    & 11.18 &   105    &&0110990301,0601010101\\
\hline
NGC278    &  13.017958  & 47.550500  & 75.5  &&    11.80 & 11.67 & 153   &2055,2056&\\
\hline
NGC300   &   13.722833 &   -37.684389  & 73.1 &      175.7     & 1.97 &    11.13 &   101 &  9883,12238 & 0112800101,0112800201,\\
&&&&&&&&&0305860301,0305860401,\\
&&&&&&&&&0656780401\\
\hline
NGC474   &   20.027878  &   3.415399&&                86.1  &     40.88  &  12.10 &   212    &&0200780101,0601670101\\
\hline
M33   &   23.462042  & 30.660222    & 1454.7&     297.1    &  0.88   &  11.24  & 109 &    786,1730,6376,6377, & 0102640101,0102640601,\\
&&&&&&&& 6378,6379,6380,6381, & 0102642301,0141980301,\\
&&&&&&&& 6382,6383,6384,6385,& 0141980501,0141980801,\\
&&&&&&&&6386,6387,6388,6389, & 0650510101,0650510201\\
&&&&&&&&7170,7171,7196,7197, &\\
&&&&&&&&7198,7199,7208,7226, &\\
&&&&&&&& 7344,7402&\\
\hline
NGC720   &   28.252077 &  -13.738653  & 178.8 &          136.9   &    23.74  &  12.51 &   289   &  492,7062,7372,8448, & 0112300101,0602010101\\
&&&&&&&&8449,11868 & \\
\hline
NGC821  &    32.088083  & 10.994917  & 188.3 &  &   23.97 & 12.15 &  220  & 5691,5692,6310,6313,&\\
&&&&&&&&6314 &\\
\hline
NGC891   &   35.639224   & 42.349146  & 171.6 &           151.6   &    10.13  &  12.19  & 227 &  794,4613,14376 &0112280101,0670950101\\
\hline
NGC1023  &    40.100042 &  39.063285  & 200.9  &    &  11.62  & 12.17 & 223  &4696, 8197,8198,8464,&\\
&&&&&&&&8465&\\
\hline
NGC1052   &   40.269994  & -8.2557642    & 64.9 &       182.6   &    19.48 &   12.15 &   220  & 5910,11355 & 0093630101,0306230101,\\
&&&&&&&&& 0553300301,0553300401\\
\hline
NGC1209   &   46.512583   & -15.611250  &&           102.5   &    31.60   & 12.24 &  235   && 0671700101\\
\hline
NGC1316   &   50.673825 &  -37.208227   &&          172.1    &   20.23  &  12.83 &  370   && 0302780101,0502070201\\
\hline
NGC1332   &   51.571884 &  -21.335216 & 74.8 &          65.9   &      19.62&    12.48 &  284 &   2915,4372 &0304190101\\
\hline
NGC1365   &   53.401548  &  -36.140402  & &          1094.4&      17.91 &   12.55&   298 && 0151370101,0151370201,\\
&&&&&&&&&0151370701,0205590301,\\
&&&&&&&&&0205590401,0505140201,\\
&&&&&&&&&0505140401,0505140501,\\
&&&&&&&&&0692840201,0692840301,\\
&&&&&&&&&0692840401,0692840501\\
\hline
NGC1386     & 54.192425  &-35.999408    & 100.5 &&      16.23 & 11.79 & 167 & 4076,12289,13185,13257&\\
\hline
NGC1395   &   54.623955 &   -23.027525    &&        61.8     &     21.46   &12.63&   318 && 0305930101\\
\hline
NGC1407    &  55.049417   & -18.580111& 59.2 &               66.9    &   23.11 &   12.92 &  396  &  7849,14033 &0404750101\\
\hline
IC342     &  56.702095 &  68.096368 & 57.8 &               202.5     & 3.35   &  11.77   &165   & 7069 & 0093640901,0206890101,\\
&&&&&&&&&0206890201,0206890401,\\
&&&&&&&&&0693850601,0693851301 \\
\hline
NGC1493    &  59.364290  & -46.210702  &&      87.4    &  11.30   & 11.42 &    126   &&0306730201,0652450101\\
\hline
NGC1521   &   62.078875   & -21.051972   & 49.4 &   140.6   &     55.64  & 12.87   & 383   & 10539 &   0503480101,0552510101\\
\hline
PGC014858 &   64.800958  &  55.874250    &&      235.4    &  64.82     & 12.08  & 208   && 0000110101,0139760101,\\
&&&&&&&&&0672050101,0672050201 \\
\hline
NGC1569  &    67.704412 &  64.847944  & 106.8   &&  2.91   & 10.97 & 89 & 782,4745&\\
\hline
NGC1600   &   67.916417   & -5.086250   & 53.5 &          94.8   &    54.70 &  13.49 &  613    & 4283,4371 & 0400490101,0400490201\\
\hline
NGC1637  &    70.367409  & -2.857962 & 121.48 &    &  10.87 & 11.57 &  141 & 766,1968,1969,1970&\\
\hline
NGC1961    &  85.519365   &69.378438    & 128.4 &      289.0  &  59* &  13.40  & 573   & 10528,10529,10530,&0673170101,0673170301,\\
&&&&&&&&10531 & 0723180101,0723180401,\\
&&&&&&&&& 0723180801,0723180201,\\
&&&&&&&&&0723180301,0723180901,\\
&&&&&&&&&0723180601,0723180501,\\
&&&&&&&&&0723180701\\
\hline
NGC2110    &  88.047420   & -7.456212  &&           59.6   &   35.6  &  12.53&   294 &&  0145670101\\
\hline
NGC2146  &    94.657125  & 78.357028 &  58.1     &  &   20.23 &   12.54&  295  & 3131,3132,3133,3134,&\\
&&&&&&&&3135,3136&\\
\hline
NGC2325   &  105.668344 & -28.697235 &&          116.8   &   23.62 &  12.04 &   203   &&0502481901,0670870101\\
\hline
NGC2276    & 111.809833  &   85.754556    & 70.3 &      54.8    &     36* &   12.05  &  203& 4968,15648 &   0022340201\\
\hline
NGC2403   &  114.214167   & 65.602556   & 217.5 &     124.2    &   3.57 &     11.43 &  127 & 2014,4627,4628,4629, &   0150651101,0150651201,\\
&&&&&&&& 4630 &0164560901 \\
\hline
NGC2681  &   133.386419 &  51.313673 & 159.9&     &  15.25 &  11.94 & 187  & 2060,2061&\\
\hline
NGC2768  &   137.906250  & 60.037222  &  64.6   &   &    20.07  &12.44 &  274 & 9528& \\
\hline
NGC2798   &  139.344970  & 41.999729  &  94.2  &  & 25.66 & 11.79 & 167 & 6729,8457,9093,10567&\\
\hline
NGC2903    & 143.042125 &    21.500833     & 93.6 &        96.3   &      9.03   & 12.03   &  200   & 11260 & 0556280301,0671430201 \\
\hline
NGC2992   &  146.425211 &  -14.326382 & 52.9 &   489.4  &    31.60  & 12.11 &  212 &3295,3296,3956 &     0147920301,0654910301,\\
&&&&&&&&&0654910401,0654910501,\\
&&&&&&&&&0654910601,0654910701,\\
&&&&&&&&&0654910801,0654910901,\\
&&&&&&&&&0654911001\\
\hline
NGC3031   &  148.888221 &    69.065295 & 319.0 &          230.5 &      3.68 &   12.14 &   219 &   735,5935,5936,5937, & 0111800101,0657801601,\\
&&&&&&&& 5938,5939,5940,5941, & 0657801801,0657802001,\\
&&&&&&&& 5942,5943,5944,5945, & 0657802201\\
&&&&&&&&5946,5947,5948,5949, & \\
&&&&&&&&9122,9805,12301& \\
\hline
NGC3079   &  150.490848 &  55.679789    &&     69.7  &        19.13   &  12.22   &232   &&0110930201,0147760101\\
\hline
NGC3115   &  151.308250  & -7.718583 & 1138.6  &&      10.17&   12.22 & 232  & 2040,11268,12095,13817,&\\
&&&&&&&&13819,13820,13821,13822,&\\
&&&&&&&&14383,14384,14419&\\
\hline
NGC3198  &   154.978966  & 45.549623   & 61.6   &&      13.87  & 11.77&  165& 9551 &\\
\hline
NGC3227   &  155.877413  &  19.865050    &&    148.0   &      20.85   &  12.14  & 217   &&0101040301,0400270101\\
\hline
NGC3312   &  159.260520 &   -27.565069  &&         68.6  &    46.06   &12.56   & 300 &&  0206230101\\
\hline
NGC3310    & 159.691083    & 53.503382  &&         142.9        &  18.10   & 11.70 &  156   &&0112810301,0556280101,\\
&&&&&&&&&0556280201\\
\hline 
NGC3379   &  161.956616  & 12.581624 & 470.4  &  &  10.46 & 12.18 &  225 & 1587,4692,7073,7074,&\\
&&&&&&&&7075,7076,11782,13829&\\
\hline
NGC3516   &  166.697876  &  72.568577     &&  518.6  &   38.9   & 12.40  & 266 &&  0107460601,0107460701,\\
&&&&&&&&&0401210401,0401210501,\\
&&&&&&&&&0401210601,0401211001\\
\hline
NGC3556   &  167.879042  & 55.674111  & 59.4&  &      13.32 & 11.95 &  189 & 2025&\\
\hline
NGC3585  &   168.321214 & -26.754840  & 94.7 & &      18.21 & 12.48 & 284& 2078,9506 &\\
\hline
NGC3608   &  169.245632   & 18.148684 &&           58.8      & 24.27   & 12.08  &  208   &&  0693300101\\
\hline
NGC3628   &  170.070710  &  13.589684& 102.4 &     65.0  &      11.30 &  12.22 &  231   & 2039,2918,2919 & 0110980101\\
\hline
NGC3631   & 170.261976 &  53.169569 &  89.1 & &     13.10 &  11.64 & 149 & 3951& \\
\hline
NGC3683  &   171.882708 &  56.877056  & 137.2  &&     33.21 & 12.11 & 214 & 4659,4660,7607&\\
\hline
NGC3690  &   172.13458  &  58.56194 &  124.9   & &  48*  & 12.84& 374 & 1641,6227,15077,15619&\\
\hline
NGC3877  &   176.532078 &  47.494346  & 101.1 &&    15.61 & 11.86  & 176 & 768,952,1971,1972&\\
\hline
NGC3898   &  177.314042  & 56.084361 & 57.4&  &   21.90 & 12.18 & 225 & 4740& \\
\hline
NGC3923   &  177.757059 &  -28.806017 & 102.1 &    150.4   &   20.88  & 12.92  & 397   & 1563,9507 & 0027340101,0602010301\\
\hline
NGC4013  &   179.630750  &  43.946583   &84.0 &       95.5   &    18.60    & 12.02 &    200  & 4013,4739 & 0306060201,0306060301,\\
\hline
NGC4039   &  180.472958 & -18.886194  &  347.6 &   267.6  &     20.88&  11.22* &    108&    3040,3041,3042,3043,&0085220101,0085220201,\\
&&&&&&&&3044,3718  & 0500070201,0500070301,\\
&&&&&&&&&0500070401,0500070501,\\
&&&&&&&&&0500070601,0500070701,\\
\hline
NGC4051  &   180.790060 &  44.531334    &&      819.0    &  14.28 &   11.82 &  171 &&  0109141401,0157560101\\
&&&&&&&&&0606320101,0606320201,\\
&&&&&&&&&0606320301,0606320401,\\
&&&&&&&&&0606321301,0606321401,\\
&&&&&&&&&0606321501,0606321601,\\
&&&&&&&&&0606321701,0606321801,\\
&&&&&&&&&0606321901,0606322001,\\
&&&&&&&&&0606322101,0606322201,\\
&&&&&&&&&0606322301\\
\hline
NGC4125   &  182.025083 &  65.174139  & 64.2  &&      24.42 &  12.90 &  390 & 2071&\\
\hline
NGC4151  &   182.635745 &  39.405730   &&        367.1   &   9.90&    11.59&   143   &&0112310101,0112830201\\
&&&&&&&&&0112830501,0143500101,\\
&&&&&&&&&0143500201,0143500301,\\
&&&&&&&&&0402660101,0402660201,\\
&&&&&&&&&0657840101,0657840201,\\
&&&&&&&&&0657840301,0657840401,\\
&&&&&&&&&0679780101,0679780201,\\
&&&&&&&&&0679780301,0679780401\\
\hline
NGC4157     &182.768208  &  50.484667 & 59.3 &       62.7  &      19.13  &  12.17&   223 &11310 &   0203170101\\
\hline
NGC4217  &   183.962083 &  47.091778   & 72.7 &&     19.58 &  12.09 & 211 & 4738&\\
\hline
NGC4244     & 184.373583 &   37.807111   &&          116.2     &   4.12  & 11.08  & 97   &&0105070201,0553880201,\\
&&&&&&&&&0553880301\\
\hline
NGC4258     & 184.739603 &  47.303973    &&     145.2  &     7.45  &  12.10 &   211   && 0059140101,0059140201,\\
&&&&&&&&&0059140401,0059140901,\\
&&&&&&&&&0110920101,0400560301\\
\hline
NGC4261   &  184.846752 &    5.825215 & 100.9 &          160.2  &    31.32   & 13.05  & 439 & 9569 &   0056340101,0502120101\\
\hline
NGC4278  &   185.028434  & 29.280756  & 578.5 &&       15.83 &  12.04&  203& 4741,7077,7078,7079,&\\
&&&&&&&&7080,7081,11269,12124&\\
\hline
NGC4291     &185.075833 &  75.370833    &&       217.5 &    33.49&    12.23  &  234   && 0124110101,0401240201,\\
&&&&&&&&&0401240301,0401240501\\
\hline
NGC4395    & 186.453592   & 33.546928  &&        154.9    &  4.49   &  10.77  &  76   &&0112521901,0112522001,\\
&&&&&&&&&0112522701,0142830101\\
\hline
NGC4414  &   186.612917  &  31.223528       &&      97.9 &      18.31    & 12.34 &  254   && 0200510101,0200510201,\\
&&&&&&&&&0402830101\\
\hline
NGC4449 &    187.046261 &  44.093630  &  100.9&   &  3.69 &  11.23 & 109 & 2031,10125,10875&\\
\hline
NGC4477   &  187.509159  & 13.636604  & 120.8 && 20.47 & 12.25  & 237 & 8066,9527,11736,12209&\\
\hline
NGC4490    &  187.650996 &   41.643898 & 97.6 &      107.9 &     8.09 &     11.58 &      142 &    1579,4725,4726 & 0112280201,0556300101,\\
&&&&&&&&&0556300201\\
\hline
NGC4507   &  188.902631 &  -39.909262     &&    138.8   & 46* & 12.40 &   267  && 0006220201,0653870201,\\
&&&&&&&&&0653870301,0653870401,\\
&&&&&&&&&0653870501,0653870601\\
\hline
NGC4565  &   189.086584  & 25.987674   &  59.2 & &       11.75 & 12.09 &  210 & 3950&\\
\hline
NGC4569    & 189.207470   & 13.162940   &&          66.0  &      12.35 &   12.06  &  206  && 0200650101 \\
\hline
NGC4593   &  189.914272 &   -5.344261  &&            115.3   &   30.80  &  12.30  &  247    &&0059830101,0109970101\\
\hline
NGC4594  &   189.997633 &  -11.623054   & 192.4 &&       10.39  & 12.89 &  388 &  1586,9532,9533&\\
\hline
NGC4618  &   190.386875 &  41.150778  &  65.9 & &7.30  & 11.25 & 110 & 7147,7098,9549&\\
\hline
NGC4631    &  190.533375 &   32.541500 & 59.2 &        54.8 &       6.32  &  11.65  &  150 & 797 &   0110900201\\
\hline
NGC4649   &  190.916564 &   11.552706  & 307.9 &    153.3 &     16.21  &   13.14 &   468 &   8182,8507,12975,12976, &  0021540401,0021540201,\\
&&&&&&&& 14328,785 & 0502160101,0502160201 \\
\hline
NGC4668   &  191.383296   & -0.535728   &&           58.2   &      16.42    & 11.22   & 107   && 0110980201\\
\hline
NGC4697    & 192.149491 &    -5.800742 &  153.8 &          65.4 &       12.31  &  12.16  &  221   & 4727,4728,4729,4730 & 0153450101\\
\hline 
NGC4736  &   192.721088  &  41.120458        &&    126.2   &   5.02  &   11.91 &   183   &&0094360601,0094360701,\\
&&&&&&&&&0404980101,0404980201\\
\hline
NGC4782   &  193.648836  & -12.568649   &&       59.4    &    54.65    & 13.78   &769 &&    0405770101,0405770201\\
\hline
PGC044532 &    194.759750  &  34.859444    &&       64.9    &   10.90   & 10.77   &76 &&  0141150101,0141150401,\\
&&&&&&&&&0141150501\\
\hline
NGC5018   &  198.254305   & -19.518193  &&     119.5  &    37.77  & 12.98  & 417   && 0502070101\\
\hline
NGC5073  &    199.836042  & -14.844528   &&           53.6&       37.39  &  12.02   &199  &&  0110980601\\
\hline
NGC5055    & 198.955542  &  42.029278    &&         68.7 &      8.29  & 12.12   &215 &&  0405080301,0405080501\\
\hline
NGC5087    & 200.104000 & -20.611000   & 49.4  &&     26.50 & 12.31  & 249 &12956& \\
\hline
NGC5128 &    201.365063 & -43.019113  &  797.4  && 3.66 &   12.09 &  211& 962,7797,7798,7799,7800,&\\
&&&&&&&&8489,8490,10722,10723,&\\
&&&&&&&& 10724,10725,11846,11847,&\\
&&&&&&&&12155,12156,13303,13304,&\\
&&&&&&&& 15294&\\
\hline
M51    & 202.469629 &   47.195172 & 854.6 &          187.7    &  7.94   & 12.15 &  220 &  354,1622,3932,12562, & 0112840201,0212480801,\\
&&&&&&&&12668,13812,13813,13814, & 0303420101,0303420201,\\
&&&&&&&& 13815,13816,15496,15553 & 0677980701,0677980801\\
\hline
IC4329    &  207.272125  &  -30.295861  &&           150.0 &       57.40  & 13.57 &  656  && 0101040401,0147440101\\
\hline
NGC5457    &  210.802267 &   54.348950 & 1106.0 &     117.6   &   6.75 &      12.00&   196    & 2065,4731,4732,4733, & 0104260101,0164560701,\\
&&&&&&&&4734,4735,4736,4737, & 0212480201\\
&&&&&&&&5297,5300,5309,5322, & \\
&&&&&&&&5323,5337,5338,5339, & \\
&&&&&&&&5340,6114,6115,6118, & \\
&&&&&&&&6152,6169,6170,6175, & \\
&&&&&&&&14341 & \\
\hline
ESO097-013  & 213.291458 &   -65.339222 & 292.4 &     168.8   &  4.21  &     11.83 &   173 &     356,365,2454,9140,& 0111240101,0701981001\\
&&&&&&&& 10937,12823,12824,& \\
\hline
NGC5506   &  213.312050   & -3.2075769      &&   298.0 &   23.83 &   12.01  &  197  && 0013140101,0013140201,\\
&&&&&&&&&0201830201,0201830301,\\
&&&&&&&&&0201830401,0201830501,\\
&&&&&&&&&0554170101,0554170201\\
\hline
NGC5529 &    213.891958 &  36.226583  &  367.0 &     &   44.13 &  12.58 &  306 &  4163,12255,12256,13118,&\\
&&&&&&&&13119&\\
\hline
NGC5643   &  218.169765  & -44.174406  &&       64.2   &   16.90  &  12.14  &  219   && 0140950101,0601420101\\
\hline
NGC5746   &  221.232992  &   1.955003  &&              341.2 &    29.04 &     13.27 &  518   && 0651890101,0651890201,\\
&&&&&&&&&0651890301,0651890401\\
\hline
PGC052940  & 222.377423  &-10.173295  & 55.0    &   72.6  & 28.8 & 11.26 & 111 & 5191 & 0149620201\\  
\hline
NGC5775  &  223.489988  &  3.544458 & 58.2  &     &    21.44 &  12.08 & 209 & 2940 &\\
\hline
NGC5813    &  225.296795   &  1.701981  & 638.2 &      172.5 &   30.15   &  12.83   & 371   & 5907,9517,12951,12952, & 0302460101,0554680201,\\
&&&&&&&& 12953,13246,13247,13253, & 055468030\\
&&&&&&&& 13255 & \\
\hline
NGC5846   &  226.622017    & 1.605625   & 149.9 &     201.0  &  26.71    & 13.03   & 431 &    788,4009,7923 & 0021540101,0723800101,\\
&&&&&&&&&0723800201\\
\hline
NGC5879  &   227.444693   & 57.000189   & 89.0 &           24.1 &      16.12   &   11.59   & 143   & 2241 &  0111260201\\
\hline
NGC5907   &   228.974042   &  56.328771     &&         142.8    &  16.24&    12.30&   247&&   0145190201,0145190101,\\
&&&&&&&&&0673920201,0673920301\\
\hline
NGC6217   &  248.163333  &  78.198167  &&       95.1  &    23.90  & 11.86   &176   && 0061940301,0061940901,\\
&&&&&&&&&0400920101,0400920201\\
\hline
NGC6300    &  259.247792  &  -62.820556  &&            46.7   &    14.43  &  12.08  & 209  &&  0059770101\\
\hline 
NGC6482  &   267.953375  &  23.071944     &&      57.7   &   58.00   &12.60 &  311 &&   0304160401,0304160501,\\
&&&&&&&&&0304160601,0304160801\\
\hline
NGC6643  &    274.943375 &   74.568361   &&           106.8  &    21.36 &    11.90  & 181   &&0602420201,0602420301,\\
&&&&&&&&&0602420401,0602420501,\\
&&&&&&&&&0602420601\\
\hline
IC4765    &  281.824695 &   -63.331331   &&           73.6   &   57.7   &13.48&  612  &&0405550401,0694610101\\
\hline
NGC6753   &  287.848500  & -57.049556      & &  73.9    &   43.60   & 13.48    &610 & &   0673170201\\
\hline
NGC6861  &   301.831167 & -48.370222  & 116.3 &   &   31.80&  12.67 & 327 & 3190,11752&\\
\hline
NGC6868  &   302.475292  & -48.379556 &  96.1  &&       31.58 &  13.01 & 425 & 3191,11753&\\
\hline
NGC6876   &  304.579792 &  -70.858806  &  75.3 &&  37.60 &  12.99 &  419 &  7059,7248&\\
\hline
NGC6946 &    308.718015 &  60.153915    &&        336.0   &  5.68   & 11.90   &  181&&   0093641701,0200670101,\\
&&&&&&&&&0200670201,0200670301,\\
&&&&&&&&&0200670401,0401360101,\\
&&&&&&&&&0401360201,0401360301,\\
&&&&&&&&&0500730101,0500730201,\\
&&&&&&&&&0691570101\\
\hline
NGC7090     & 324.120250 &  -54.557333 & 56.7 &              49.4     &   8.55  &  11.42 &   125 &  7060,7252 & 0200230201,0503460101\\
\hline
NGC7213   &  332.317958  &  -47.166611  & &      182.1  &    22.00  & 12.57  &  304   && 0111810101,0605800301\\
\hline
NGC7176   &  330.535178   & -31.989762  & 49.5 &     134.1 &   34.86 &  12.58 &     307   & 905 & 0147920601,0202860101,\\
&&&&&&&&&0414580101\\
\hline
ESO467-051  & 335.819208  &  -28.980972      && 74.5    &   19.65 &  10.57* & 67   &&0691590101,0700381701\\
\hline
NGC7320  &   339.014083  &  33.948111 &112.9&      46.0    &   15.67   & 11.21  &  107  & 789,7924 &0021140201\\
\hline
IC5264    &  344.221000 & -36.554167  &   58.8  &  &  47.75  &  12.11 & 214 &  2196& \\
\hline
IC5267    &  344.306542  & -43.396139    &55.0 &     67.3   &     26.08   &12.51  & 290 &3947 &  0306080101\\
\hline
NGC7552   &  349.044830   &-42.584742     & &  105.8 &     17.15 & 11.96 &  190 & &  0501650201,0501650301,\\
&&&&&&&&&0093640701,0093641401\\
\hline
NGC7582  &   349.597917 &  -42.370556     &&         170.0   &   20.94 &    12.30&   246   &&0112310201,0204610101,\\
&&&&&&&&&0405380701\\
\hline
NGC7619 &    350.060549 &   8.206246  & 64.2 &  &    53.31 & 13.43 & 589 & 2074,3955&\\
\hline
NGC7673  &   351.920913 &  23.589039  &  58.7 &  &   49*  &  11.72 & 158 & 9554&\\
\hline
IC5332   &   353.614542  & -36.101083 &  107.4 &      &   8.40 & 11.30 & 115 & 2066,2067 &\\
\hline
NGC7714    & 354.058744  & 2.1551615  &59.0  &  &     30.92 &11.71& 156& 4800& \\
\hline
NGC7796    & 359.749057 &  -55.458316 & 72.3 &           84.2 &      49.43   & 13.04 &   436 &    7061,7401 &0693190101\\
\enddata
\tablecomments{Galaxies and observations examined in this analysis. The columns $t_{\text{CXO}}$ and $t_{\text{XMM}}$ are the total integration times with these telescopes for each galaxy. The distance $d$ is the redshift-independent average distance listed at NED; galaxies without redshift-independent estimates have their distances marked with asterisks, and for them we estimate the distance from the redshift assuming WMAP7 cosmology \citep{Komatsu2011}. We estimate $M_{\text{halo}}$ and $R_{\text{vir}}$ from the 2MASS K-band total magnitude, as discussed in section 2.1.3 (galaxies without 2MASS K-band magnitudes have these values noted with asterisks, and K-band magnitudes from other surveys are used instead). }
\end{deluxetable}

\end{document}